\documentclass[journal,10pt]{IEEEtran}
\usepackage{algorithmic}
\usepackage{algorithm}
\usepackage{color}
\usepackage{subfigure}
\usepackage{amsmath,graphicx}
\usepackage{multirow}
\usepackage{csquotes}
\usepackage{array}
\usepackage{makecell}
\usepackage{graphicx}
\usepackage{amssymb}
\usepackage{enumitem}
\DeclareMathOperator*{\argmin}{argmin}

\newcommand*{\argminl}{\argmin\limits}

\usepackage{verbatim} 
\usepackage{cite}

\usepackage{booktabs,graphicx,makecell,siunitx,eqparbox}

\newcommand{\bigO}{\mathcal{O}}
\usepackage{balance}
\usepackage{scalerel}
\usepackage{tikz}
\usetikzlibrary{svg.path}

\definecolor{orcidlogocol}{HTML}{A6CE39}
\tikzset{
  orcidlogo/.pic={
    \fill[orcidlogocol] svg{M256,128c0,70.7-57.3,128-128,128C57.3,256,0,198.7,0,128C0,57.3,57.3,0,128,0C198.7,0,256,57.3,256,128z};
    \fill[white] svg{M86.3,186.2H70.9V79.1h15.4v48.4V186.2z}
                 svg{M108.9,79.1h41.6c39.6,0,57,28.3,57,53.6c0,27.5-21.5,53.6-56.8,53.6h-41.8V79.1z M124.3,172.4h24.5c34.9,0,42.9-26.5,42.9-39.7c0-21.5-13.7-39.7-43.7-39.7h-23.7V172.4z}
                 svg{M88.7,56.8c0,5.5-4.5,10.1-10.1,10.1c-5.6,0-10.1-4.6-10.1-10.1c0-5.6,4.5-10.1,10.1-10.1C84.2,46.7,88.7,51.3,88.7,56.8z};
  }
}

\newcommand\orcidicon[1]{\href{https://orcid.org/#1}{\mbox{\scalerel*{
\begin{tikzpicture}[yscale=-1,transform shape]
\pic{orcidlogo};
\end{tikzpicture}
}{|}}}}

\usepackage{hyperref} 

\begin{document}

\title{Sparsifying Dictionary Learning for Beamspace Channel Representation and Estimation in Millimeter-Wave Massive MIMO}

\author{Mehmet Ali~Ayg\"{u}l \orcidicon{0000-0002-1797-8238}\,,~\IEEEmembership{Student Member,~IEEE,}
Mahmoud~Nazzal \orcidicon{0000-0003-3375-0310}\,, and~H\"{u}seyin~Arslan \orcidicon{0000-0001-9474-7372}\,,~\IEEEmembership{Fellow,~IEEE}

\thanks{This work was supported by the Scientific and Technological Research Council of Turkey (TÜBİTAK) under grant No. 116E078.}
\thanks{M. A.~Ayg\"{u}l is with the Department of Electronics and Communications Engineering, Istanbul Technical University, Istanbul, 34467, Turkey and also with the Department of Research \& Development, Vestel, Manisa, 45030, Turkey (e-mail: mehmetali.aygul@ieee.org).
}
\thanks{M.~Nazzal is with the Department of Electrical and Electronics Engineering, Istanbul Medipol University, Istanbul, 34810, Turkey (e-mail: mahmoud.nazzal@ieee.org).
}
\thanks{H.~Arslan is with the Department of Electrical Engineering, University of South Florida, Tampa, FL, 33620, USA and also with the Department of Electrical and Electronics Engineering, Istanbul Medipol University, Istanbul, 34810, Turkey (e-mail: arslan@usf.edu).}
\thanks{This work has been submitted to the IEEE for possible publication. Copyright may be transferred without notice, after which this version may no longer be accessible.}
}

\maketitle 
\begin{abstract}
Millimeter-wave (mmWave) massive multiple-input-multiple-output (mMIMO) is reported as a key enabler in the fifth-generation communication and beyond. It is customary to use a lens antenna array to transform a mmWave mMIMO channel into a beamspace where the channel exhibits sparsity. This beamspace transformation is equivalent to performing a Fourier transformation of the channel. Still, a Fourier transformation is not necessarily the optimal one, due to many reasons. Accordingly, this paper proposes using a learned sparsifying dictionary as the transformation operator leading to another beamspace. Since the dictionary is obtained by training over actual channel measurements, this transformation is shown to yield two immediate advantages. First is enhancing channel sparsity, thereby leading to more efficient pilot reduction. Second is improving the channel representation quality, and thus reducing the underlying power leakage phenomenon. Consequently, this allows for both improved channel estimation and facilitated beam selection in mmWave mMIMO. Besides, a learned dictionary is also used as the precoding operator for the same reasons. Extensive simulations under various operating scenarios and environments validate the added benefits of using learned dictionaries in improving the channel estimation quality and the beam selectivity, thereby improving the spectral efficiency.
\end{abstract}

\vspace{-0.2cm}

\begin{IEEEkeywords}
Antenna arrays, artificial intelligence, millimeter wave radio propagation
\end{IEEEkeywords}

\IEEEpeerreviewmaketitle
\vspace{-0.3cm}

\section{Introduction}
\label{Section1}
\vspace{-0.1cm}

\par \IEEEPARstart{M}{assive} multiple-input-multiple-output (mMIMO) is widely considered as a key enabler for wireless communication in the era of the fifth generation and beyond. This is because of its ability to improve the system data rate \cite{nassar}. Especially, when it operates at millimeter-wave (mmWave) frequencies, it has crucial importance. This allows for increased data rates due to the higher spectral efficiency \cite{6375940} and wider bandwidth \cite{5783993}. However, the main challenge with mmWave mMIMO is the hardware and power requirements.

\par Beamforming techniques are used to reduce the cost and power consumption by suppressing the co-channel interference and improving the signal-to-noise ratio (SNR) at the receiver end \cite{kutty2015beamforming}. These techniques can be divided into three groups; analog, digital, and hybrid. Analog beamforming is cost and power-effective but only supports one data-stream at a time \cite{xiao2017millimeter}. On the other hand, digital precoding uses a radio frequency (RF) chain per antenna element and thus requires high power consumption, complexity, and cost. Therefore, a hybrid precoding technique has been introduced as a compromise to both settings \cite{DFT3}.

\par Hybrid precoding connects hundreds of antennas to a small number of RF chains through analog phase shifters \cite{PR}. However, realizing mmWave mMIMO is still a non-trivial task since the numbers of antennas \cite{Lsa} and RF chains \cite{MA} are still high. On the other hand, mMIMO channels show strongly directional propagation with low dimensionality properties at mmWave frequencies. This motivates a beamspace representation \cite{heath2016overview} where channel sparsity can be exposed. This sparsity can be exploited with the advent of compressive sensing (CS), allowing for reduced channel training and feedback overheads.

\vspace{-0.2cm}

\subsection{Related Works and Motivation}

\par CS-based channel estimation algorithms exploit angle domain sparsity of mmWave mMIMO channels \cite{12n,15n,16n}. CS allows for sub-Nyquist sampling by enabling sparse signal recovery at a sampling rate below the Nyquist rate. However, these algorithms are designed with high-resolution phase shifters for hybrid precoding systems. Still, a phase shifter network can be replaced by a lens antenna array (LAA) \cite{sayeed2010continuous} for a further reduction in the hardware cost and power consumption. Hence, an LAA is widely used to expose a beamspace channel representation in mmWave mMIMO. Therefore, the dimension of a mmWave mMIMO channel can be reduced by beam selection over the sparse beamspace channel \cite{zeng2016millimeter,xie2019power}.

\par A promising channel estimation technique for the case of using an LAA is sparsity mask detection \cite{SMD2}. In this setting, the beams of large power are determined initially. Then, the dimension of the beamspace channel is reduced and it is estimated in this reduced dimension. However, scanning over all the beams is a time-consuming process. Another algorithm to reduce the number of antennas is the support detection (SD) algorithm for sparse coding. This algorithm divides the channel estimation problem into a series of subproblems, each of which only considers one channel path component \cite{Dai_Journal}. To this end, this multitude of beamspace channel estimation algorithms models the impact of the LAA by a discrete Fourier transform (DFT) matrix. DFT discretizes the continuous angular channel parameter space into a finite set of predefined spatial angles. This set covers the whole angular beam range and emphasizes sparsity. Thus, the performance of these algorithms largely depends on how accurately this discretization can model the true sparsity of the channel, i.e., it depends on the representation power of this sparsifying basis/transform.

\par
\begin{figure}[t]
\centering
\resizebox{0.97\columnwidth}{!}{
\includegraphics[width=14cm]{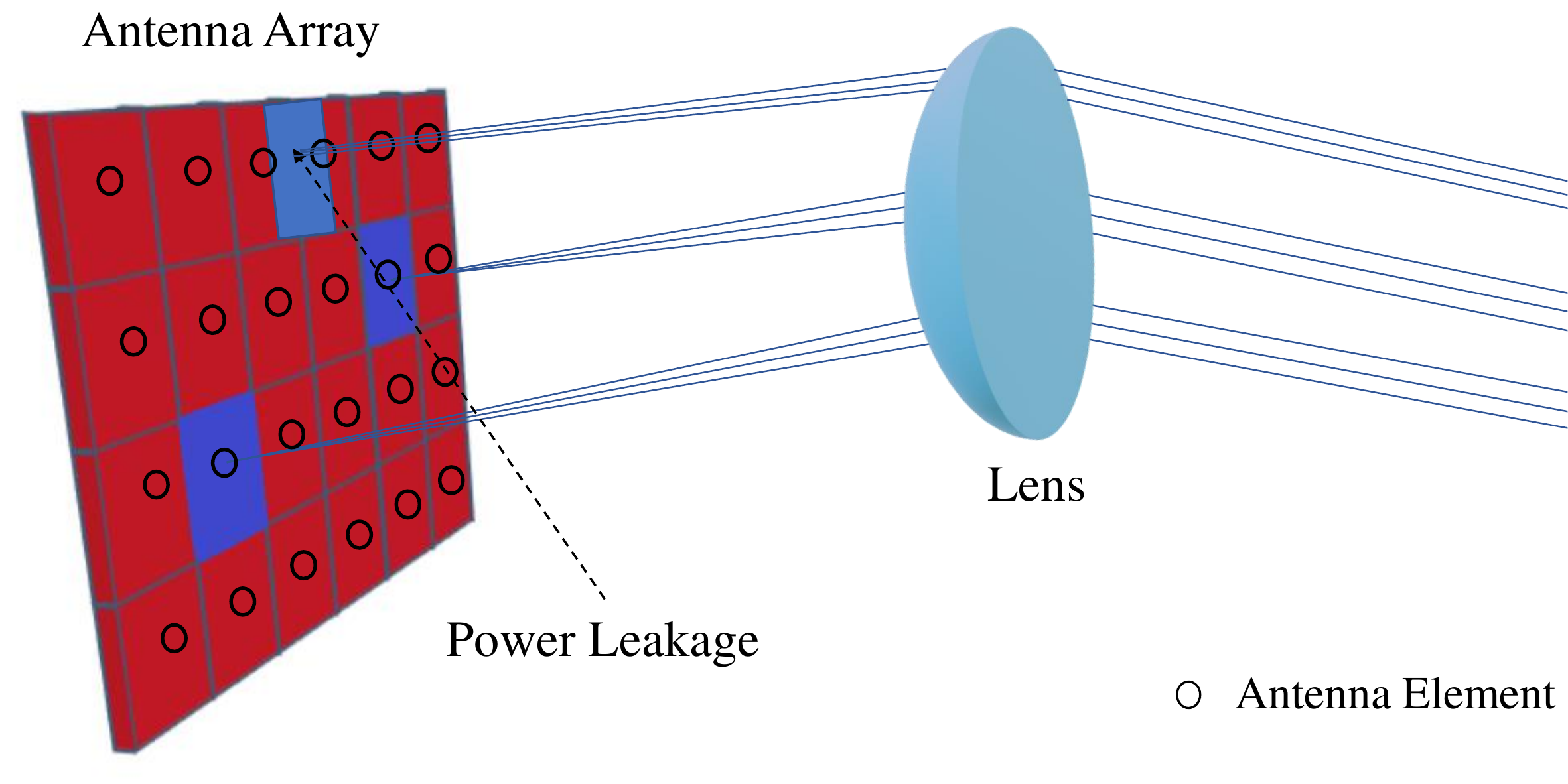}}
\caption{The concept of power leakage in LAAs.}
\label{intro}
\end{figure}

 Despite achieving the state-of-the-art performance in mmWave mMIMO channel estimation, a DFT sparsifying basis is known to have several inherent shortcomings. Specifically, the actual angles of departure (AoDs) of paths are continuously distributed since the spatial sampling points of the LAA are not finite and fixed in practice. Therefore, the AoD of a path will not necessarily match the spatial sample points of the LAA \cite{tang2013compressed} modeled by DFT, as illustrated in Fig.~\ref{intro}. Consequently, the power of a beam will leak onto multiple beams in the beamspace (\textit{off-grid} problem) \cite{sayeed2010continuous}. This \textit{power leakage} effect is serious even for the simplest cases and incurs obvious SNR losses \cite{xie2019power}. \cite{he2018deep,dong2019deep,wei2020deep} estimate a beamspace channel using a machine learning framework to improve estimation quality. However, these works do not address the sparsity of the representation. Therefore, their performance improvement is limited.

\par The problem of obtaining an efficient sparsifying transform is studied in the context of signal representation. It has been shown that one can use a \textit{redundant} (over-complete) DFT basis in pursuit of achieving a finer discretization of the channel signal space. Along with this line, a redundant dictionary is tailored for LAA to combat the power leakage caused by the continuous angles of multipath components in \cite{wan2019compressive}. However, there are certain limits for the redundancy of this basis, as it substantially increases the computational cost. Besides, high redundancy creates the side effect of more similarity between the columns of the basis matrix, thereby degrading the representation quality. Therefore, recent research considers using a limited degree of redundancy in the sparsifying basis, while trying to tackle the off-grid effects. Although \cite{tropp2007signal, gurbuz2018sparse, tang2018off, SMD2} use moderate degrees of redundancy, their computational complexity is still prohibitively large.

\par Rather than expanding the quantity of discretized points, recent literature calls for developing new beamspace transformation operators to combat off-grid effects. Such operators are not restricted to having the DFT character. For example, the Fourier domain is shown to overlook the Dirichlet structure inherent in mmWave channels \cite{guvenc}. Thus, the authors proposed using a set of Dirichlet kernels to serve as a sparsifying dictionary. Besides, the DFT is shown to be sub-optimal as a sparsifying transform \cite{gwon2012compressive}. So, the authors proposed using a Karhunen-Lo\`eve transform (a.k.a. principal component analysis) as a data-dependent optimal basis. Alternatively, enhanced and more general dictionaries are anticipated to offer a better alternative to DFT bases \cite{heath2016overview}. A dictionary that is generated by a finer-grained point further improves the approximation of the continuous points and also the estimation quality \cite{berger2010application}. In this context, dictionary learning offers dictionaries with enhanced sparsity and representation quality leading to lowering the severity of power leakage, more efficient pilot reduction, and improved channel estimation quality.

\vspace{-0.4cm}

\subsection{Contributions, Notation, and Organization}

\par In view of the above discussion, a preliminary version of this work \cite{our} showing the advantage of using a learned dictionary for precoding in mmWave mMIMO is extended. This paper proposes an algorithm for beamspace channel representation and estimation based on sparse coding over a learned dictionary. Here is a brief account of the contributions in this paper.
\begin{itemize}[leftmargin=*]
\item \textit{Using a learned dictionary as the channel sparsifying transform operator:} As opposed to standard beamspace channel sparsification, such as using the DFT to represent LAA operation, a learned dictionary enhances the channel sparsity. This leads to a further reduction in hardware, cost, and power consumption. Besides, this allows for easier beam selection at the receiver end. Along this line, the usage of a learned dictionary that is obtained by training over previous channel observations is proposed as the channel sparsifying transform operator.
\item \textit{Dictionary learning for precoding:} Usage of a different learned dictionary is also proposed as the precoding operator for the same reasons. Such a dictionary is obtained by training over example precoding matrix realizations.
\end{itemize}

\par Extensive simulations under various operating scenarios and environments validate that using learned dictionaries for beamspace channel sparsification and precoding improves the channel estimation quality and enhances the beam selectivity by improving the spectral efficiency. This improvement is especially the case when the antenna array is not exactly uniform.

\par \textit{Notation:} Plain-faced letters represent scalars. Bold-faced lower-case and bold-faced upper-case letters denote vectors and matrices, respectively. In a matrix $\boldsymbol{X}$, the symbol $\boldsymbol{X}_i$ represents its $i$th column. Similarly, $\boldsymbol{x}_i$ is the $i$th element in a vector $\boldsymbol{x}$. The Hermitian conjugate transpose symbol is denoted by $\dagger$. $\boldsymbol I_K$ is the $K \times K$ identity matrix. The ${\|.\|}_2$, $\|.\|_0$ and $tr$ symbols represent the 2-norm, the number of nonzero elements in a vector, and the trace operation, respectively.

\par \textit{Organization:} This paper is organized as follows. Section~\ref{Section2} revises the preliminaries and presents the system model. The proposed algorithm for channel representation and estimation is detailed in Section~\ref{Section3}. Section~\ref{Section4} presents experiments conducted to evaluate the performance of the proposed algorithm, and the paper is concluded in Section~\ref{Section5}.

\section{Preliminaries and System Model}
\label{Section2}

\subsection{Dictionary Learning for Sparse Recovery}

\par Let $\boldsymbol{D} \in \mathbb{C}^{N \times K}$ denote a sparsifying transform operator (dictionary). A signal $\boldsymbol{y} \in \mathbb{C}^N$ is said to have a sparse representation in $\boldsymbol{D}$ if it can be approximated as $\boldsymbol{y}\approx\boldsymbol{Dw}$. Here, $\boldsymbol{w}\in\mathbb{C}^K$ denotes a sparse coding coefficient vector composed mainly of zeros. For a given $\boldsymbol{y}$ and $\boldsymbol{D}$, $\boldsymbol{w}$ can be obtained through the following sparse recovery process.
\begin{equation}
\operatorname*{arg\,min}_{\boldsymbol{w}} {\|\boldsymbol{w}\|}_0 ~ s.t. ~ {\|\boldsymbol{y}-\boldsymbol{Dw}\|}_2^2 <\epsilon,\label{eq1}
\end{equation}
\noindent where $\epsilon$ is an error tolerance.

\par It is noted that the problem in (\ref{eq1}) is NP-hard as one has to solve for the positions and magnitudes of the nonzero elements in $\boldsymbol{w}$. Still, there are two main approaches to approximately solve this problem. The first approach is the family of greedy pursuit algorithms that offer efficient approximate solutions by iteratively minimizing the number of nonzeros in $\boldsymbol{w}$. Second is the $\ell_{1}$-relaxation approach that relax $\ell_{0}$ to the $\ell_{1}$ norm. This relaxation offers a loose bound on sparsity but achieves a significant reduction in the computational cost. A benchmark sparse representation technique is the orthogonal matching pursuit (OMP) \cite{OMP}.

\par A sparsifying dictionary represents the transformation matrix to a domain in which the signal of interest is sparse. To this end, there are two main families of dictionaries. First is mathematically-defined basis functions, such as the DFT and discrete-cosine transform matrix. These are easy to prepare. However, they may not necessarily transform into the domain that exhibits signal sparsity. Second is learned dictionaries. A learned dictionary, especially if redundant, promotes sparsity, enhances the representation quality, and is locally adaptive to the signals of interest. In essence, this dictionary is composed of prototype signals as its columns. These signals are rich in structure as compared to fixed basis vectors.

\par In the learned dictionary, one learns a dictionary by training over a set of example training signals $\boldsymbol{Y} \in \mathbb{C}^{N\times M}$ through an artificial intelligence procedure referred to as \textit{dictionary learning}, described as follows
\begin{equation}
\operatorname*{arg\,min}_{\boldsymbol{W,D}}{\|\boldsymbol{W}_i\|}_0 ~ s.t. ~ {\|\boldsymbol{Y}_i- \boldsymbol{DW}_i\|}_2^2 < \epsilon ~\forall~ ~i.\label{2}
\end{equation}

\par The K-SVD algorithm \cite{ksvd} is one of the widely used algorithms for a dictionary learning process. In this algorithm, first, the parameter $\Lambda_i$ of nonzero elements of the $i$-th row of $\boldsymbol{W}$ is determined for each dictionary atom $\boldsymbol{D}_i$. Then, a partial residual matrix is calculated and its columns are restricted to the active set of signals that use the $i$-th atom for their sparse approximation. Finally, the atom $\boldsymbol{D}_i$ and the coefficients $\boldsymbol{W}_{{\Lambda}_i}^{i}$ are updated using the solution of the best rank-1 approximation of the matrix, which can be calculated using its SVD. More explanation can be found in \cite{nazzal2015structural}.

\subsection{Why The Learned Dictionary Is Better Than DFT?}

\par A DFT basis is essentially a mathematically defined basis function where its basis vectors are defined to uniformly quantize the directions in the vector space. Thus, it is a generic basis, and the success of its representation directly depends on the extent to which a given signal is aligned to the (fixed) basis functions that span the directionality in the vector space. Conversely, a learned dictionary has learned vectors as its columns. These vectors are trainable parameters over a comprehensive set of example signals in a machine learning operation referred to as the dictionary learning/training process. Therefore, each dictionary vector forms a prototype signal, and it is thus commonly referred to as an atom. Hence, dictionary atoms are obtained by learning over training data rather than uniformly sampling the space based on a certain criterion. This learning enjoys the generalization properties of machine learning, i.e., a learned dictionary is expected to work well with new and unforeseen data points. Therefore, this inherent data-fitting property empowers dictionaries to better represent signals of the same class of its training set more sparsely and compactly, as opposed to generic bases like the DFT. In essence, signals of any type may belong to a specific subspace and may not necessarily be spread all over the vector space. Therefore, a custom-made basis like the dictionary is tailored to best fit data of a certain type, for example, images, channel responses, or beams.

\begin{figure}
\centering
\begin{tabular}{@{}c@{}}
\includegraphics[width=.78\linewidth]{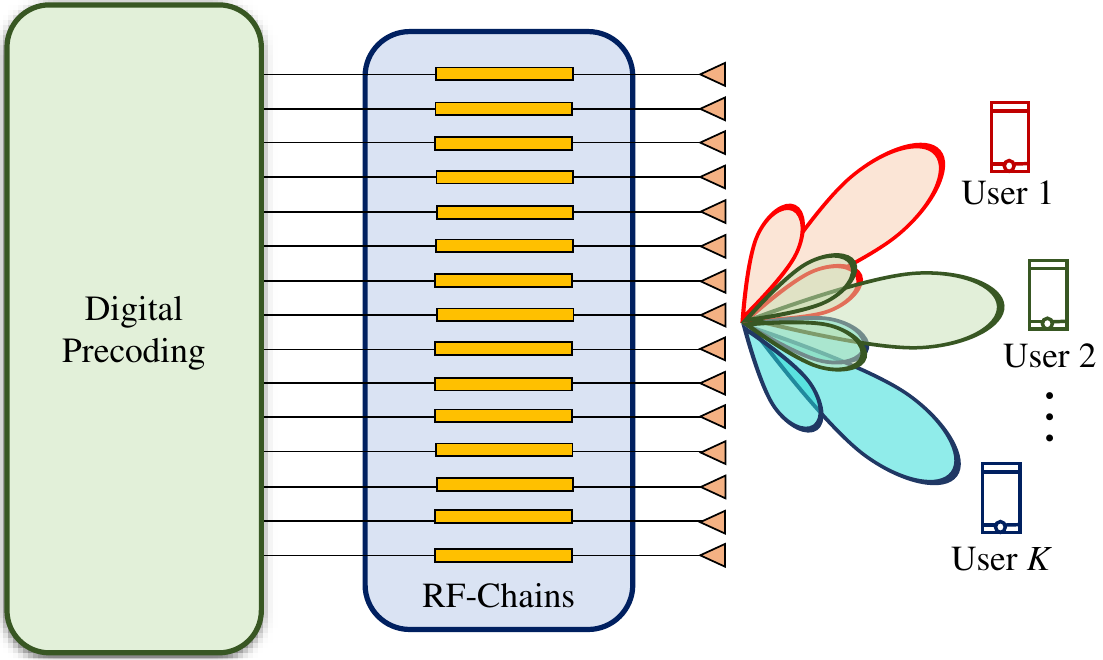} \\
\small (a) 
\end{tabular}
\begin{tabular}{@{}c@{}}
\includegraphics[width=.99\linewidth]{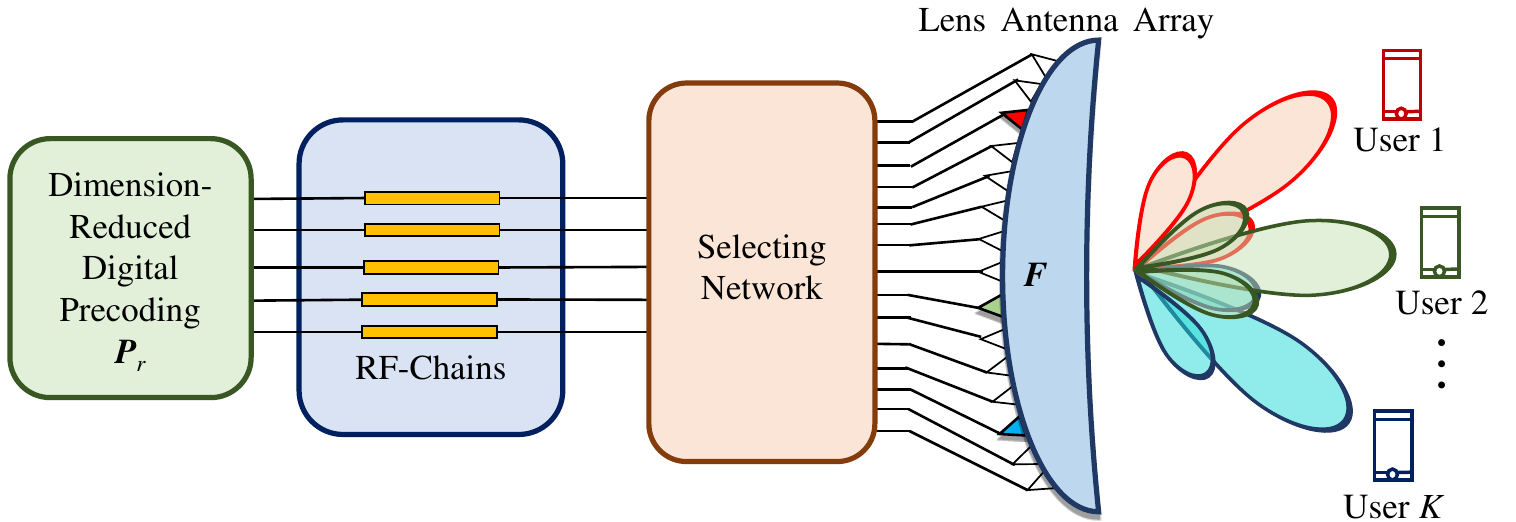} \\
\small (b) 
\end{tabular}
\caption{Antenna array configuration in mMIMO; (a) conventional and (b) with an LAA controlled by sparse coding beam selection \cite{Dai_Journal}.}
\label{MIMO}
\end{figure}

\subsection{System Model}

\par This paper considers a mmWave mMIMO system running in time division duplexing (TDD). The base station (BS) uses $N$ antennas with $N_{\text{RF}}$ RF chains to serve $K$ single-antenna users (UEs). Figure~\ref{MIMO} (a) shows a conventional mmWave mMIMO setting. The $K\times1$ received signal vector $\boldsymbol{y}_{\text{DL}}$ of all $K$ UEs in the downlink (DL) for the conventional MIMO systems in the spatial domain can be presented as
\begin{equation}
\boldsymbol y_{\text{DL}}=\boldsymbol H^{\dagger} \boldsymbol {Ps}+\boldsymbol n,\label{3}
\end{equation}
\noindent where the DL channel matrix is denoted by $\boldsymbol H^{\dagger}\in\mathbb{C}^{K\times N}$, $\boldsymbol H=[\boldsymbol{h}_1,\boldsymbol{h}_2,\cdots,\boldsymbol{h}_K]$ is the uplink (UL) channel matrix according to the channel reciprocity \cite{15n}, $\boldsymbol h_k$ of size $N\times1$ is the channel between the $k$th the UE and the BS, $\boldsymbol s$ of size $K\times1$ is the data signal vector for all $K$ UE with normalized power $\mathbb{E}(\boldsymbol {ss}^\dagger) = \boldsymbol I_K$, $\boldsymbol P \sim N\times K$ is the precoding matrix. This matrix satisfies the total transmit power constraint $\rho$ as $tr \mathbb{(\boldsymbol P \boldsymbol P^\dagger}) \le \rho$. Finally, $\boldsymbol n \sim \mathcal{CN}\mathbb(0,\sigma_{\text{DL}}^{2}\boldsymbol I_K)$ is the $K\times1$ additive white Gaussian noise vector, where $\sigma_{\text{DL}}^{2}$ is the DL noise power. Figure~\ref{MIMO} (a) shows that in conventional MIMO systems, the number of RF chains needed is equal to the number of antennas. i.e., $N_{\text{RF}} = N$, which is mostly large for mmWave mMIMO systems, e.g., $N_{\text{RF}}= N=$ 256 \cite{Lsa}.

\par Two channel models are used in the paper; the Saleh-Valenzuela (SV) and the geometry-based stochastic channel model (GSCM). Despite their similarity, SV model is primitive and widely used in mmWave channel modeling whereas GSCM better reflects the operation of antenna arrays and it can form the benchmark for mMIMO channel modeling, as a more advanced model \cite{nazzal2020channel}. Therefore, we opted to use both channel models to represent mmWave mMIMO.

\subsubsection{The Saleh-Valenzuela Channel Model}

\par The SV channel model is customarily used to model mmWave channels as it accounts for their low-rank nature. According to this model, the channel is expressed as follows \cite{Mmwave,PR}
\begin{equation}
\boldsymbol{h}_{k}=\sqrt{\frac{N}{L+1}}\sum_{i=0}^{L}\beta_{k}^{(i)}\boldsymbol{a}\left(\psi_{k}^{(i)}\right)=\sqrt{\frac{N}{L+1}}\sum_{i=0}^{L}\boldsymbol{c}_{i},\label{eq4}
\end{equation}
\noindent where the line-of-sight (LoS) component of the $k$th UE is $\boldsymbol c_{0}=\beta_{k}^{(0)}\boldsymbol{a}(\psi_{k}^{(0)})$. Also, $\beta_{k}^{(0)}$ represents the complex gain and $\psi_{k}^{(0)}$ denotes the spatial direction. The non-LoS (NLoS) component of the $k$th UE is $c_i=\beta_{k}^{(i)}\boldsymbol{a}(\psi_{k}^{(i)})$ for $1\le i\le L$ and the total number of NLoS components, denoted by $L$, is usually obtained by channel measurement \cite{ME}. Besides, $\boldsymbol{a}(\psi)$ is the $N\times1$ array steering vector. For a typical linear array with $N$ antennas, the steering vector can be represented as follows \cite{18}
\begin{equation}
\boldsymbol{a}(\theta)=\frac{1}{\sqrt{N}}[1,\ e^{-j2\pi \psi_i(\theta)}, \ldots, e^{-j2\pi \psi_i(\theta)} (N-1)]^{\dagger}, \label{eq5}
\end{equation}
\noindent where the direction of physical propagation is denoted by $\theta$ and the spatial direction is defined as $\psi_i \triangleq \frac{d_i}{\lambda}\sin(\theta)$ \cite{Mmwave}, $\lambda$ denotes the wavelength, and $d_i$ represents the antenna spacing in the $i$th column and it is usually $\lambda/2$ for linear antenna array.

\subsubsection{Geometry-Based Stochastic Channel Model}

\par The GSCM is also used as it is a more realistic channel model. For this model, the DL channel vector is considered from the BS to the $k$th UE. This can be represented as \cite{Ding_Rao}
\begin{equation} 
{\boldsymbol{h}}_k = \sum _{i=1}^{N_c} \sum _{l=1}^{N_s} \beta _{k}^{(i,l)} {\boldsymbol{a}}(\theta {_k}^{(i,l)}),\label{eq6}
\end{equation}
\noindent where the complex gain of the $l$th scattering cluster is denoted by $\beta^{i,l}$, the number of scattering clusters is denoted by $N_c$, and the number of sub-paths per scattering cluster is denoted by $N_s$. The symbol $\theta^{i,l}$ denotes the angle-of-arrival/ angle-of-departure (AoA/AoD) of the $l$th subpath in the $i$th scattering cluster. The steering vector $\boldsymbol{a}(\theta^{(i,l)})$ represents the normalized array response at the UE.

\par
\begin{figure}[t]
\centering
\resizebox{0.6\columnwidth}{!}{
\includegraphics[width=14cm]{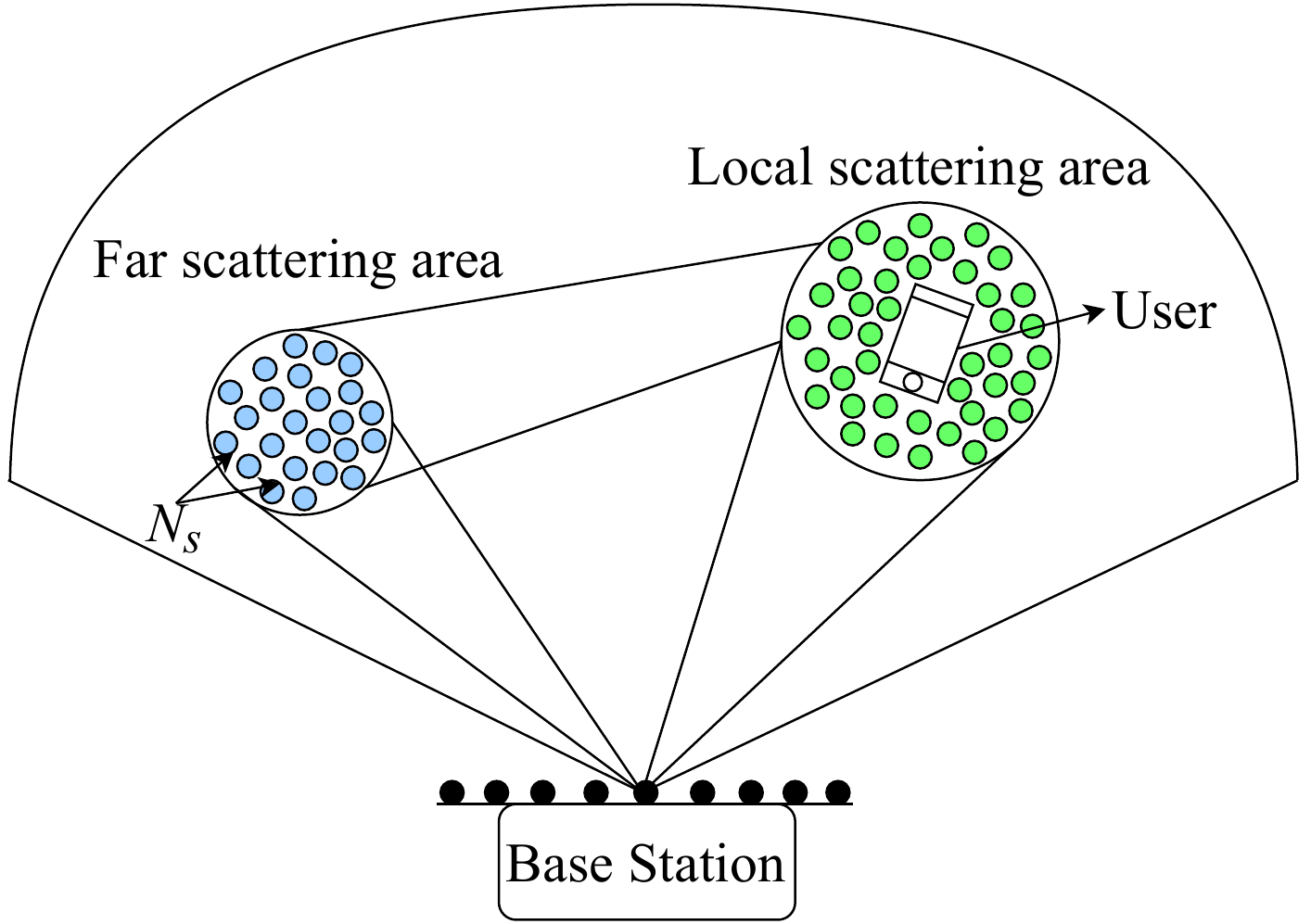}}
\caption{The GSCM concept \cite{GSCM}. In this configuration, local scatterers are centred around the UE and far scatterers are far away from both UE and BS.}
\label{GSCM}
\end{figure}

 For scattering, the principles of GSCM are adopted as in Fig.~\ref{GSCM}. In this figure, far scatterers represent mountains, high-rise buildings, etc. Also, they determine the locations of the dominant scattering clusters for a specific cell and are common
to all the users irrespective of user position. We assume that these are far away from the BS. Thus, the subpaths associated with a specific scattering cluster will be concentrated in a small range, i.e., having a small angular spread. While modeling the scattering effects that are UE-location dependent (e.g., the ground reflection close to the UE or some moving physical scatterers near the UE), we assume the UE is far from the BS. Thus, subpaths associated with the UE-location-dependent scattering cluster also have a small angular spread. Since the BS is far away and is commonly assumed to be mounted at a height, the number of scattering clusters contributing to the channel responses is limited, i.e., $N_c$ is small.

\subsection{MmWave mMIMO Channels in the Beamspace}

\par Transforming the conventional channel \cite{MA} to a beamspace representation can be done conveniently using an LAA \cite{Mmwave}, as demonstrated in Fig.~\ref{MIMO} (b). Particularly, a well-designed LAA plays the role of a spatial DFT matrix $\boldsymbol U$ that comprises the array steering vectors of $N$ orthogonal directions (beams) covering the entire angle space. This matrix can be represented as \cite{Mmwave}
\begin{equation}
\boldsymbol{U}=[\boldsymbol{a}(\bar{\psi}_{1}),\boldsymbol{a}(\bar{\psi}_{2}),\cdots,\boldsymbol{a}(\bar{\psi}_{N})]^{\dagger}, \label{8}
\end{equation}
\noindent where $\bar{\psi}_n= \frac{1}{N}(n-\frac{N+1}{2} )$ for $n=1,2, \dots , N$ are previously defined spatial directions by LAA. Then, the system model of mmWave mMIMO with an LAA can be represented by
\begin{equation}
\tilde{\boldsymbol{y}}^{\text{DL}}=\boldsymbol{H}^{\dagger}\boldsymbol{U}^{\dagger}\boldsymbol{BP}_{r}\boldsymbol{s}+\boldsymbol{n}=\tilde{\boldsymbol{H}}^{\dagger}\boldsymbol{BP}_{r}\boldsymbol{s}+\boldsymbol{n}, \label{9}
\end{equation}
\noindent where the received DL signal in the beamspace is
$\tilde{\boldsymbol{y}}_{\text{DL}}$, $\boldsymbol{\tilde{H}}^\dagger=\boldsymbol{H}^{\dagger}\boldsymbol{U}^{\dagger}=\boldsymbol{(UH)}^\dagger$ represents the DL beamspace channel matrix, in which $N$ columns being $N$ orthogonal beams, $\boldsymbol{B}$ of size $N\times K$ form the selecting matrix whose entries belong to $\{0, 1\}$. As an example, when the $n$th beam is selected by the $k$th UE, the element of $\boldsymbol{B}$ at the $n$th row and the $k$th column would be 1. After that, $\boldsymbol{P}_r$ of size $K\times K$ is the dimension-reduced digital precoding matrix.

\par It should be noted that by the virtue of the limited number of dominant scatters in the mmWave prorogation environments \cite{Lsa}, a beamspace channel $\boldsymbol{\tilde{H}}^\dagger$ (or evenly $\boldsymbol{\tilde{H}}$) has a sparse structure \cite{Mmwave,Kuser}. Consequently, it is obvious from Fig.~\ref{MIMO} (b) that a small number of beams can be selected to decrease the effective channel dimension, without causing an evident wastage in the performance. Moreover, a small number of RF chains is needed since a small-size digital precoder $\boldsymbol{P}_r$ is needed. In practice, however, it is challenging to obtain a beamspace channel in a large size with a limited number of RF chains. Specifically, the channel dimension is large while the number of RF chains is limited, and the signals on all antennas cannot be sampled simultaneously.

\subsection{Pilot Transmission}

\par In this paper, all of the UEs transmit pilot sequences to the BS over $Q$ instants to estimate the beamspace channel. Also, the beamspace channel remains unchanged within such channel coherence time as in \cite{3}, $Q$ instants are divided into $M$ blocks, and each block consists of $K$ instants such as $MK^2$. For the $m$th block, $\boldsymbol{\Psi}$ pilot matrix with a size of $K\times K$ is used as in \cite{Dai_Journal}. Then, according to the channel reciprocity in TDD systems \cite{15n} the received UL signal matrix can be represented as

\begin{equation} \boldsymbol{\tilde{Y}}_{m}=\boldsymbol{\tilde{H}}\boldsymbol{\Psi}_m+\boldsymbol{N}_m,\label{new1} \end{equation}
\noindent where $m=1, 2, \ldots, M$ and $\boldsymbol{N}_m$ is the $N\times K$ noise matrix in the $m$th block.

\section{Learned Dictionaries for Beamspace Channel Representation and Estimation}
\label{Section3}

\subsection{Power Leakage in Beamspace Channels}

\par The AoDs in an mMIMO system are distributed continuously in the angular domain. However, modeling the lens operator with a DFT basis limits the angular spread to include specific sample points. Thus, an AoD of a specific propagation path should not necessarily be matched by the given sample points. This causes the power of a path to leak onto multiple beams in the beamspace channel \cite{Mmwave}, as known as power leakage \cite{xie2019power}. For a single-UE single-path scenario, when a uniform linear array (ULA) is used, the worst power leakage is \cite{xie2019power}
\begin{equation} \eta_{ULA}=1-\frac{1}{2\sum\nolimits_{i=1}^{N/2}\frac{\sin^{2}(\pi/2N)}{\sin^{2}((2i-1)\pi/2N)}}. \label{nula} \end{equation}
\noindent With the system models considered in this paper, the worst power leakage is around 0.60, according to (\ref{nula}), which is quite high.

\begin{figure}[t]
\centering
\resizebox{0.67\columnwidth}{!}{
\includegraphics[width=14cm]{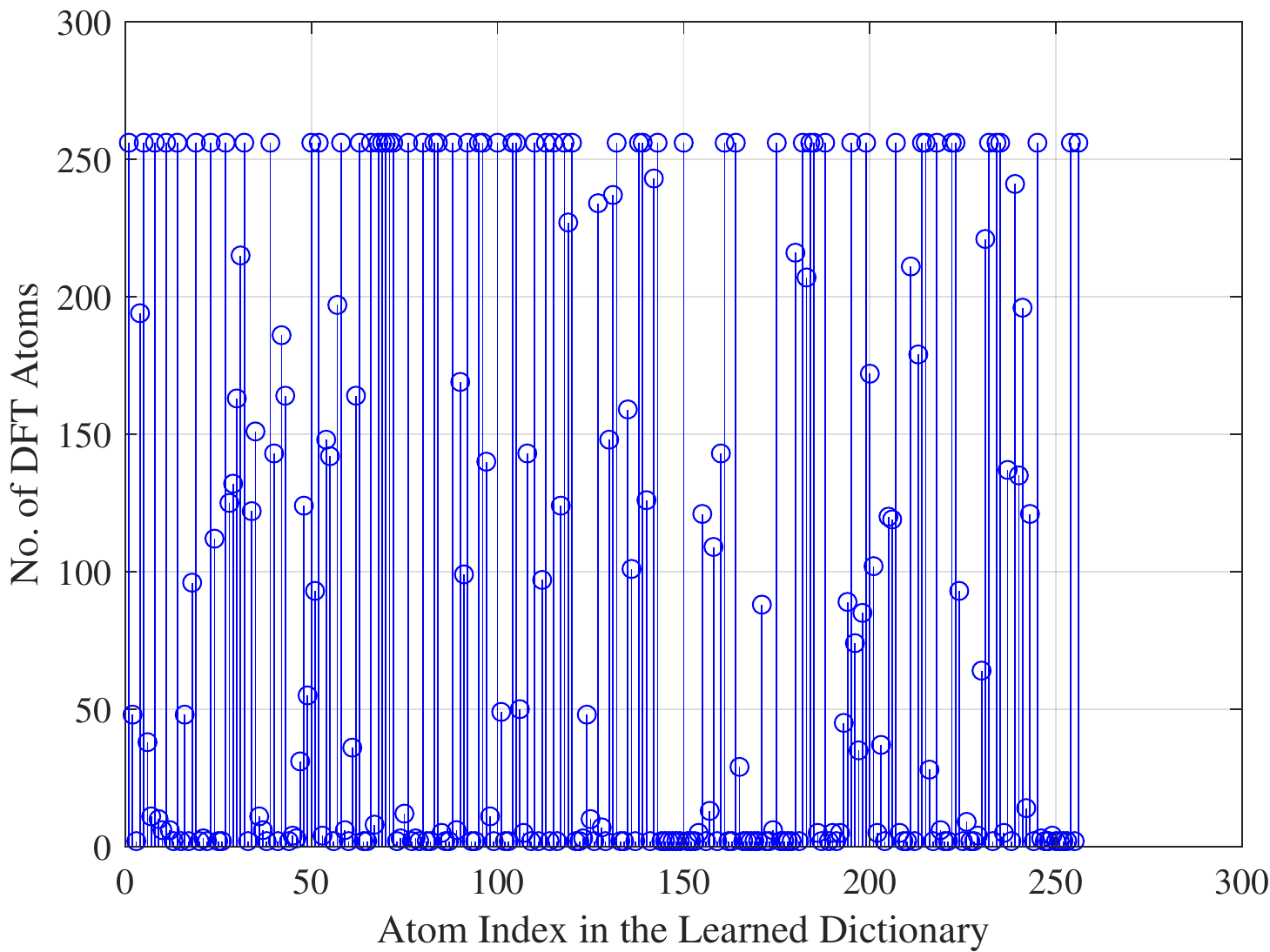}}
\caption{Beamspace sampling: a dictionary beam corresponds to a composite of DFT-modeled beams.}
\label{mot1}
\end{figure}

\begin{figure}[!t]
\centering
\begin{tabular}{@{}c@{}}
\includegraphics[width=.76\linewidth]{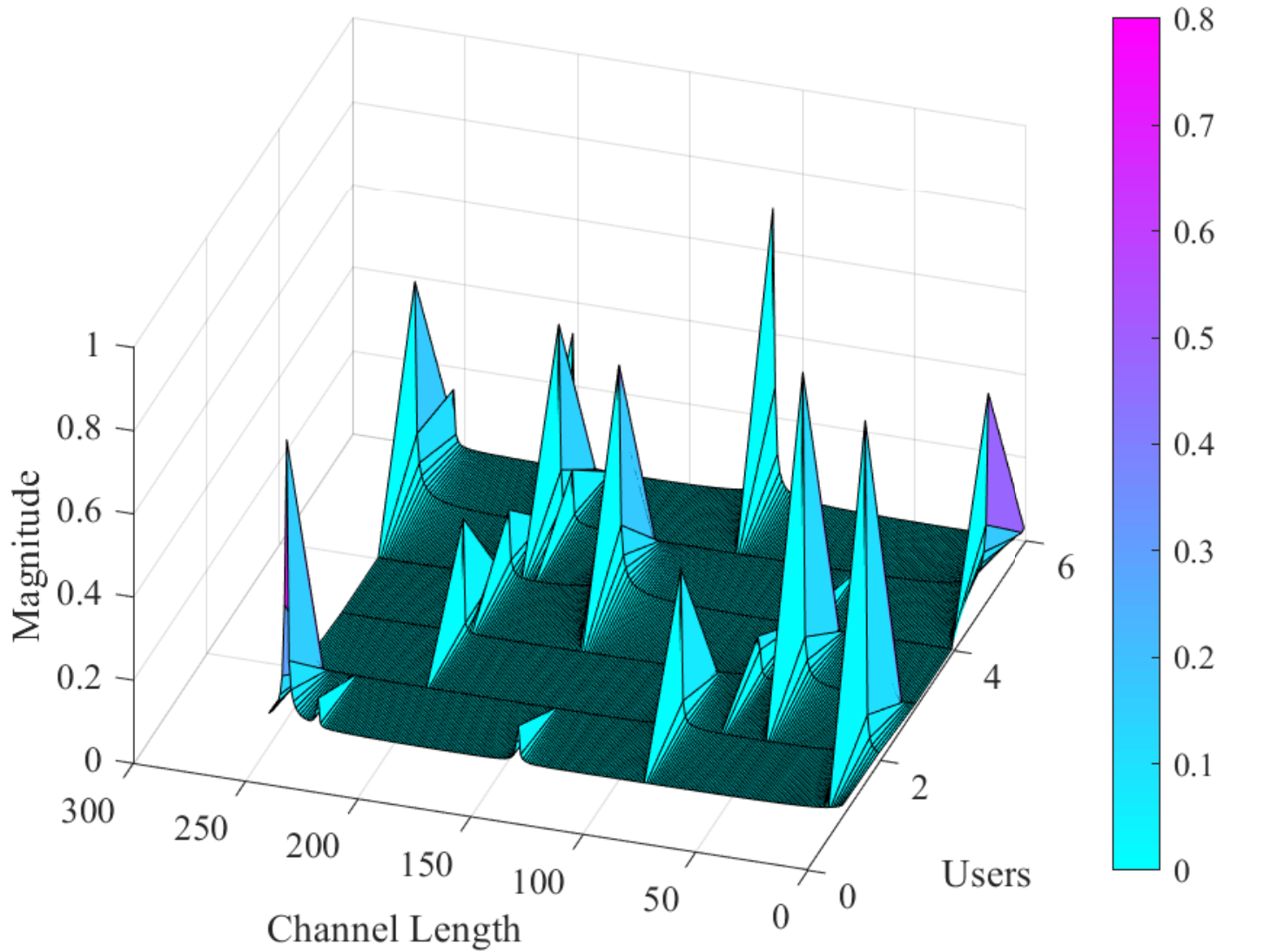} \\
\small (a) 
\end{tabular}
\begin{tabular}{c}
\includegraphics[width=.76\linewidth]{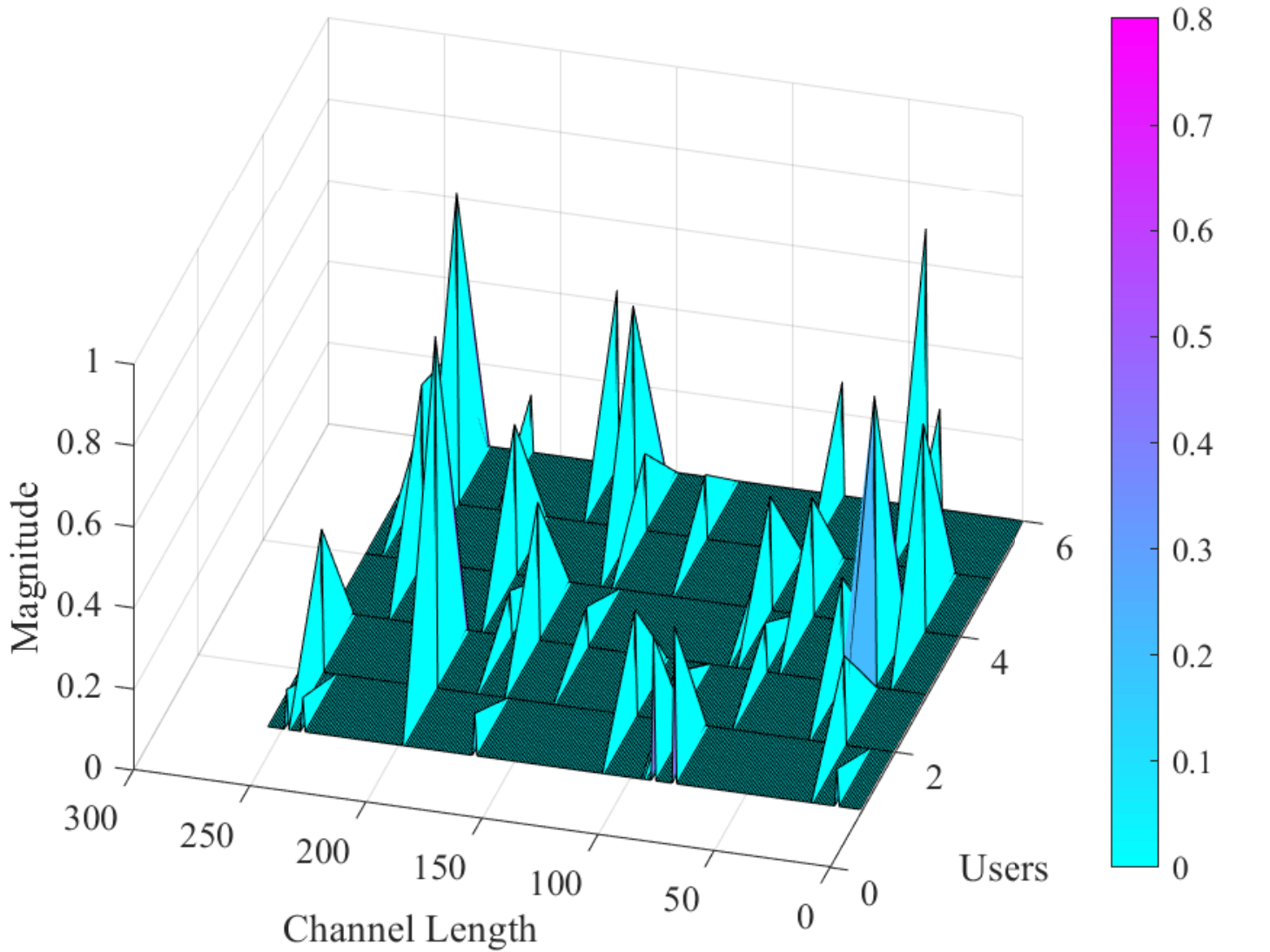} \\
\small (b) 
\end{tabular}
\caption{Magnitudes of beamspace channel coefficients obtained with (a) the DFT and (b) a learned dictionary for a multiple-UEs multiple-paths scenario.}
\label{2Dmotivation}
\end{figure}

\subsection{Motivating Examples}

\par Beamspace channel sparsity is a key channel estimating prior. However, due to the power leakage and the many nonzero elements, a beamspace channel is not ideally sparse \cite{7454701}. Therefore, using a better sparsifying transformation allows for revealing the sparsity of the channel in a better fashion. As a preliminary example of this idea, Fig.~\ref{mot1} illustrates the number of DFT columns that are necessary to represent the signal within 90\% of its energy. The figure reveals that a dictionary atom can be viewed as a composition of multiple DFT columns. In other words, each dictionary atom is rich as structure compared to a DFT column vector.

\par
\begin{figure}[!t]
\centering
\begin{tabular}{@{}c@{}}
\includegraphics[width=.56\linewidth]{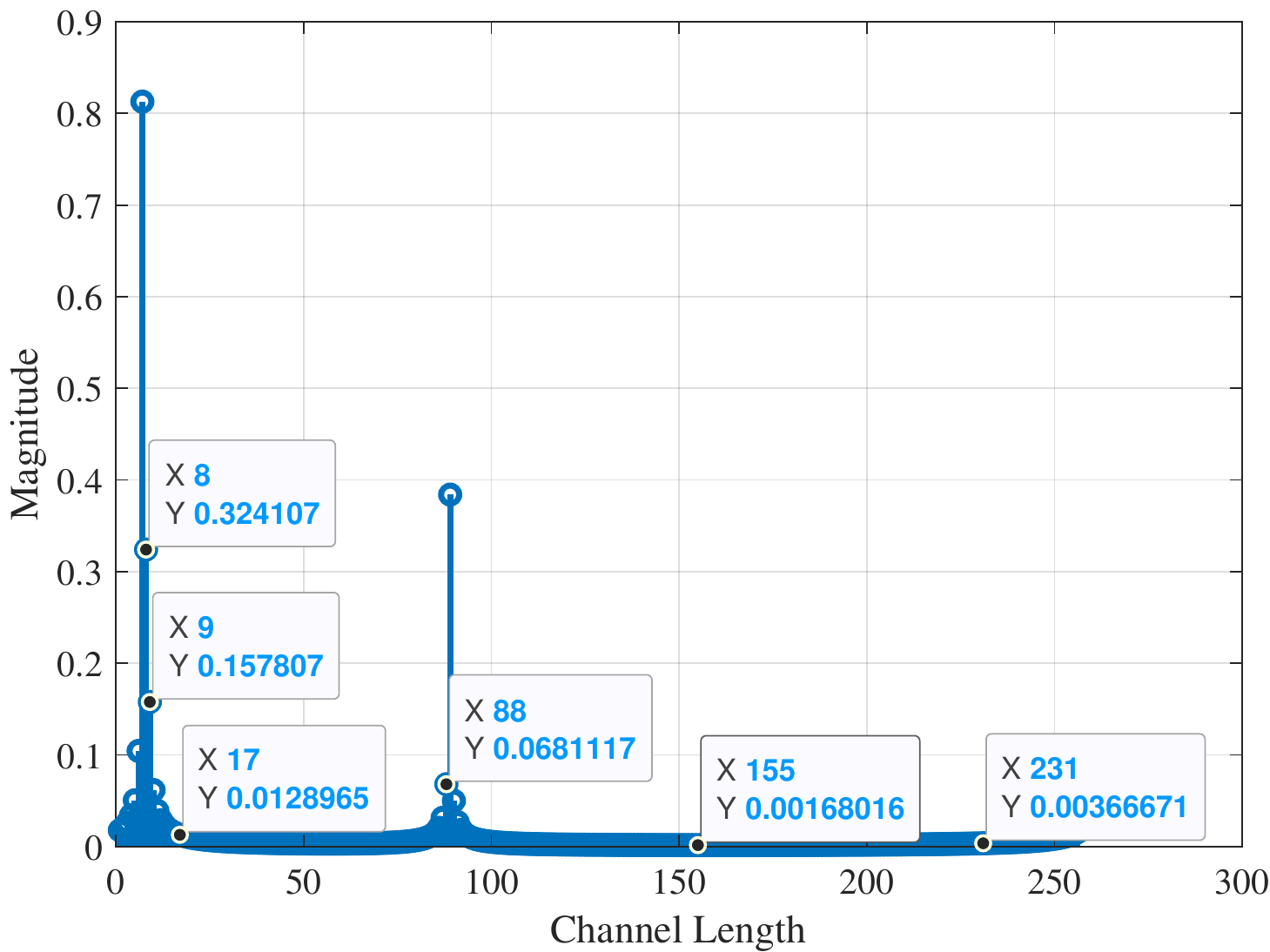} \\
\small (a) 
\end{tabular}
\begin{tabular}{c}
\includegraphics[width=.56\linewidth]{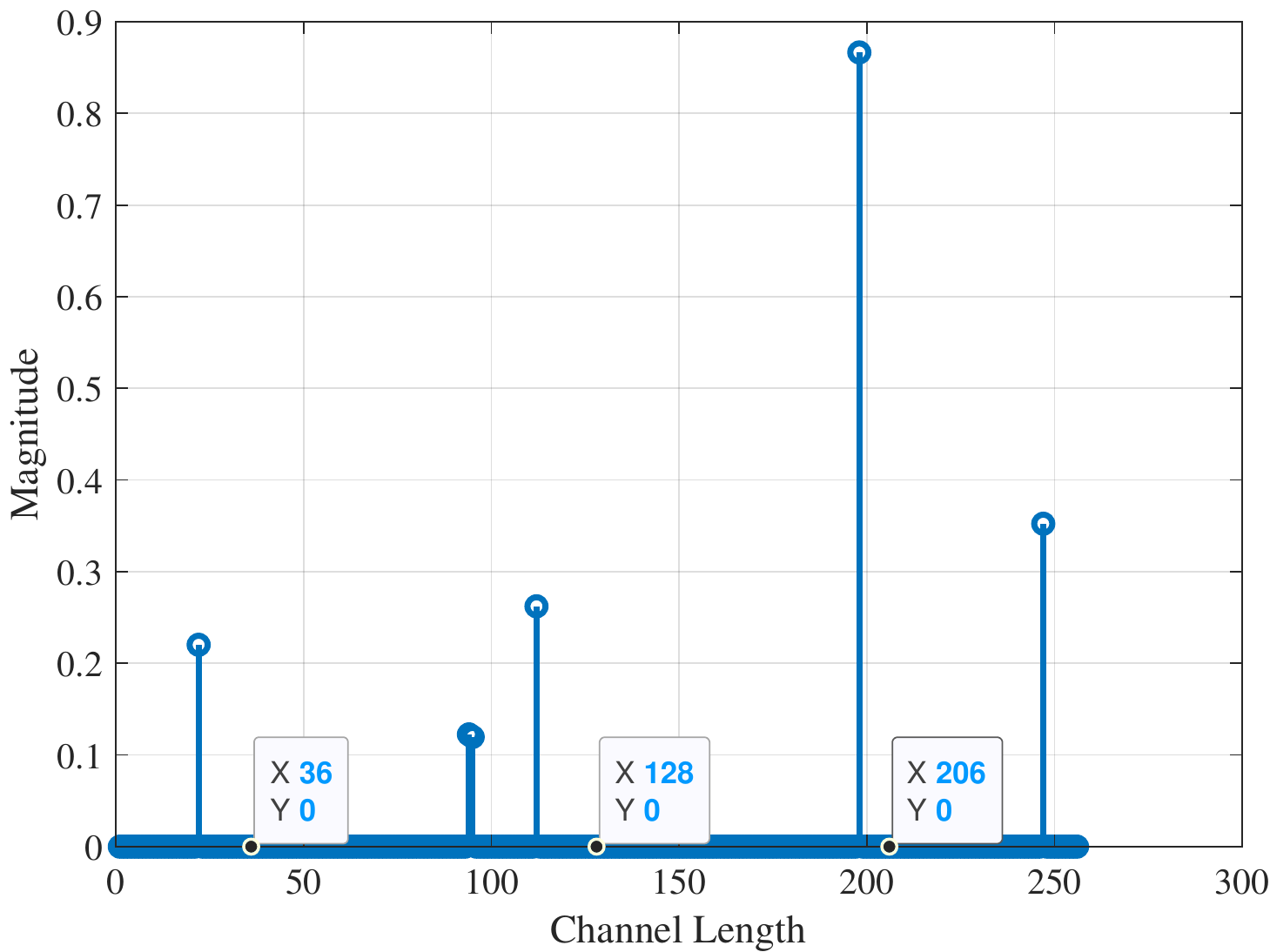} \\
\small (b) 
\end{tabular}
\caption{Magnitudes of beamspace channel coefficients obtained with (a) the DFT and (b) a learned dictionary for a a single-UE multiple-path scenario.}
\label{1Dmotivation}
\end{figure}

 The above-mentioned empirical conclusion is further illustrated in Fig.~\ref{2Dmotivation}. This figure shows the magnitudes of beamspace channel coefficients obtained by the DFT resembling the space defined by using an LAA and a learned dictionary for multiple-UEs multiple-path scenarios. In both cases, the dictionary is obtained by training it over a set of channel realizations. Any standard dictionary learning algorithm can be used for this purpose, such as the K-SVD algorithm used in this paper. Besides, the dictionary size is $96\times 256$. Also, the SV channel model is used where it has four multiple-path components and is generated according to the specification presented in Section~\ref{Section4}. It can be seen from Fig.~\ref{2Dmotivation} that DFT magnitudes exhibit side lobes around the nonzero elements which are the smaller shapes that are just next to the main shapes. Besides them, even far elements from the main lobes are nonzero. On the other, far elements from the main lobes are zero and there are no side lobes in the dictionary learning-based algorithm. These are evident beamspace sparsity is enhanced in the space defined by the dictionary. On the contrary, one that was created with dictionary learning does not have such a thing. This is further illustrated in Fig.~\ref{1Dmotivation}, where analyses are made for a single UE. The figure clearly shows that the DFT-based channel has side lobes and the dictionary learning-based channel is more sparse. The improved sparsification obtained with the learned dictionary is intuitively expected to improve the channel estimation quality. This proposition is analytically investigated in the Appendix.

\vspace{-0.2cm}

\subsection{The Proposed Algorithm}

\par In view of the above-mentioned motivations, this paper proposes to use the learned dictionaries for channel representation and precoding.

\begin{figure}[!t]
\centering
\begin{tabular}{c}
\includegraphics[width=.98\linewidth]{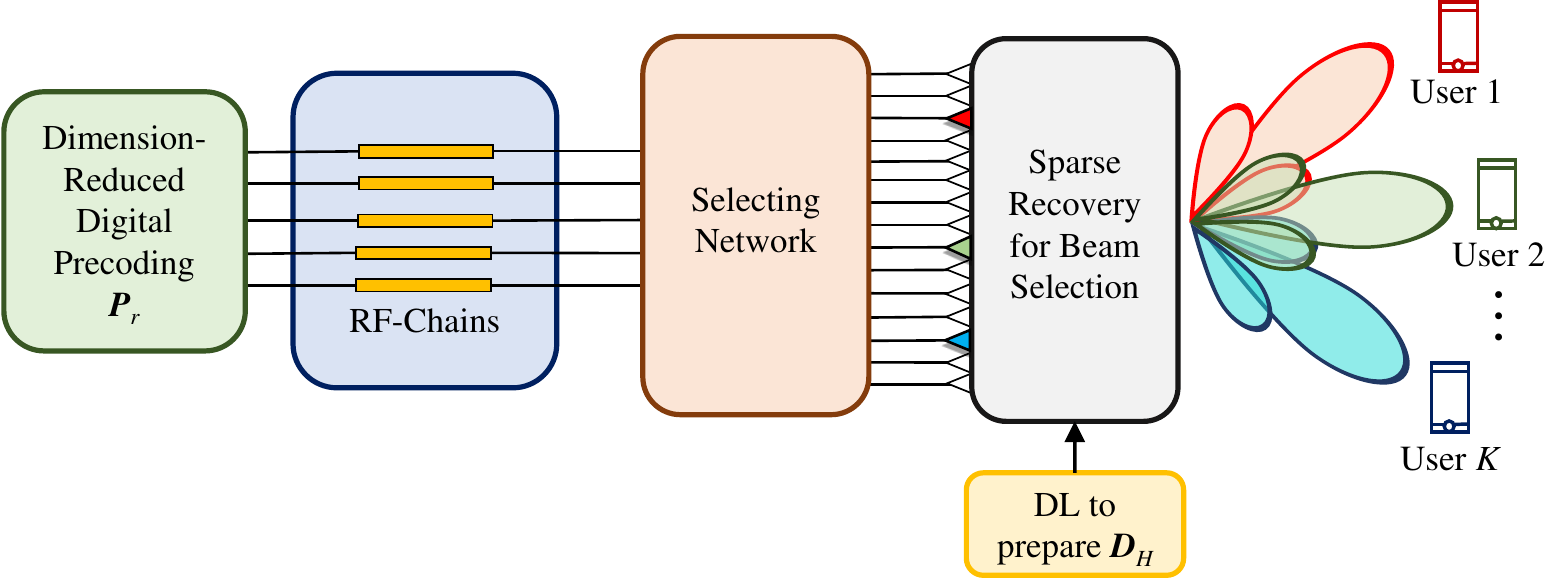} \\
\small (a) 
\end{tabular}
\begin{tabular}{@{}c@{}}
\includegraphics[width=.98\linewidth]{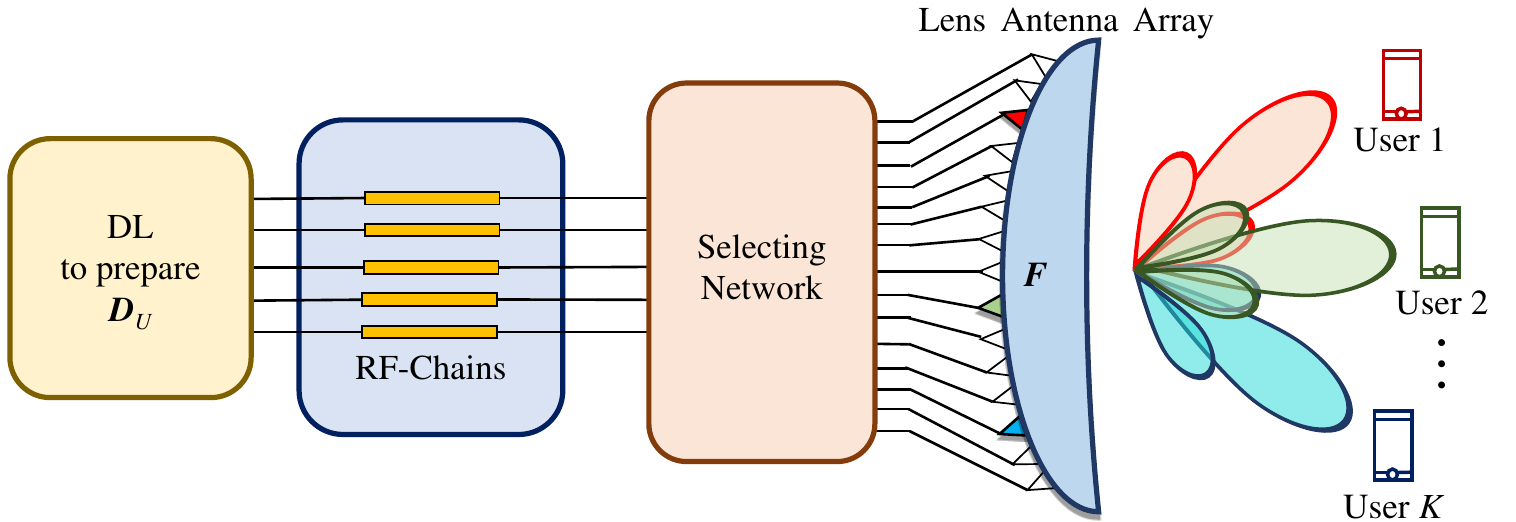} \\
\small (b) 
\end{tabular}
\begin{tabular}{c}
\includegraphics[width=.98\linewidth]{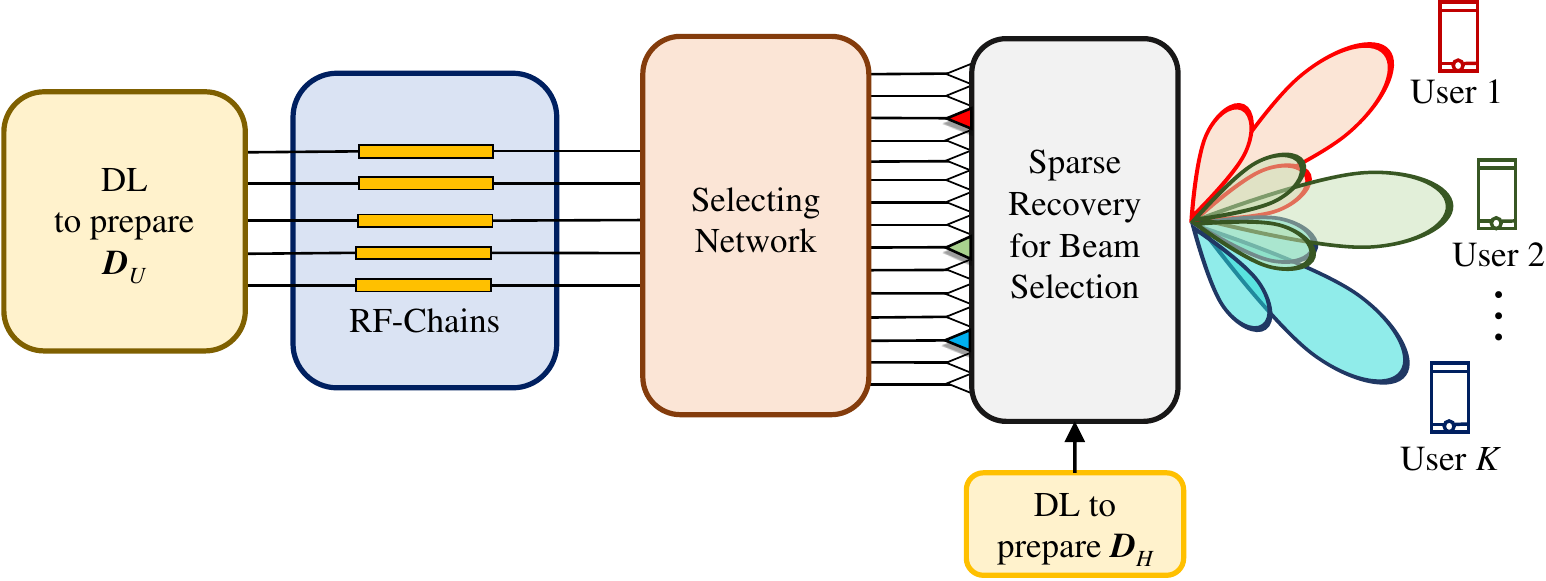} \\
\small (c) 
\end{tabular}
\caption{Antenna array configuration in mMIMO when (a) a channel representation is made by a learned dictionary, (b) a sparse coding beam selection controlled by a learned dictionary, and (c) a channel representation and a sparse coding beam selection made by learned dictionaries.}
\label{MIMOjournal}
\end{figure}

\subsubsection{The Proposed Algorithm for Beamspace Channel Representation}

\par This algorithm consists of training and testing stages. In the training stage, a set of UL channel realizations is obtained to learn a dictionary. Here note that a set of channel realizations can be obtained by classical channel estimation algorithms in the literature \cite{hassan2020channel}. In these techniques training signal is sent from the receiver and its response is observed at the transmitter end\footnote{Based on the channel reciprocity training signal also can be sent by the transmitter and observed at the receiver but in that case, the channel information should be shared with the transmitter as well since the dictionary learning will be done at the transmitter.}. It is noted that this process will be done periodically (for example, every night) by the BS to learn any far scatterer changes\footnote{Channel measurements of the signals reflected from the same far scatterers contain signals with similar incident angles \cite{DFT1,molisch2014propagation}. In fact, local scattering changes do not affect the representation of the dictionary. This is because machine learning algorithms (e.g. a dictionary learning algorithm) are powerful for denoising \cite{kaur2018review} and so they reduce the effect of local scatterers.} in the environment.

\par
\begin{figure}[bht!]
\centering
\resizebox{0.85\columnwidth}{!}{
\includegraphics[width=14cm]{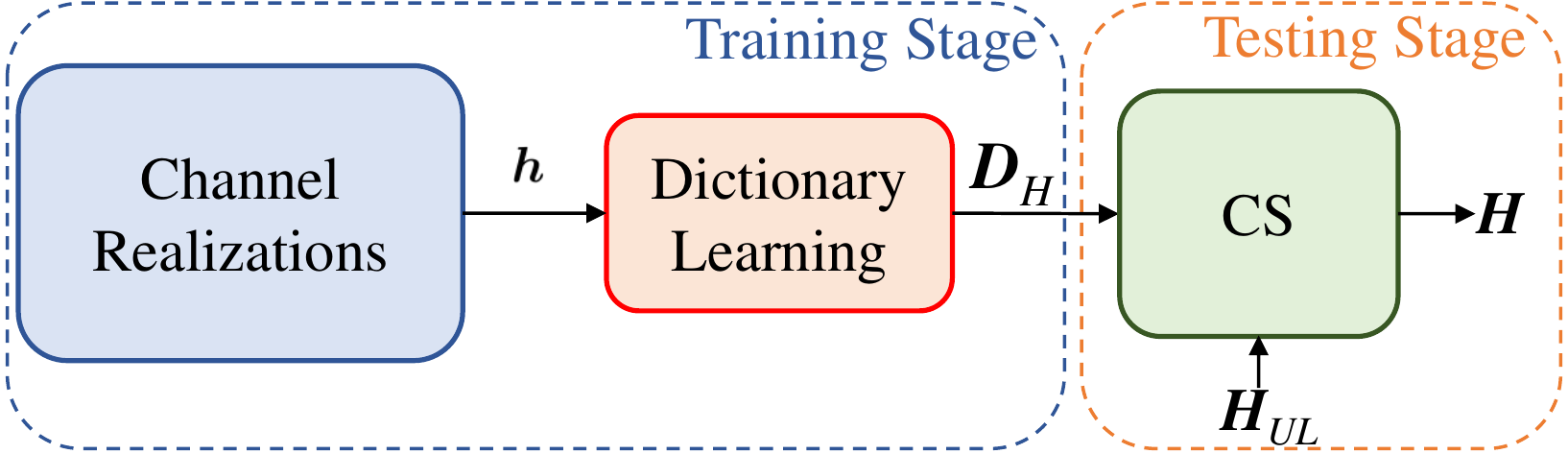}}
\caption{The diagram of the proposed dictionary learning-based algorithm to convert the channel to beamspace channel.}
\label{Diagram1}
\end{figure}

 After a set of channel realizations is obtained, a dictionary is trained over this set. A learned dictionary corresponds to using composite (multiple) lenses. In other words, a learned dictionary is represented by many DFT basis functions, as detailed in the Appendix. In the testing stage, a CS algorithm is applied with the learned dictionary ($\boldsymbol{D}_H$) and UL channel ($\boldsymbol{H}_{UL}$)\footnote{The channel between the receiver and transmitter can be measured at the UL training mode in mmWave mMIMO TDD systems \cite{sun2019beam}.} to generate a beamspace channel ($\boldsymbol{H}$). Here note that the UL channel only represents the environment (without the effect of the lens antenna array and dictionary learning). This environment can be learned with classical channel estimation algorithms \cite{arslan2001channel}. Then, a beamspace channel can be created based on the learning environment and a CS algorithm. The block diagram of these processes is represented in Fig.~\ref{Diagram1}. Besides, the proposed algorithm for channel representation is illustrated in Fig.~\ref{MIMOjournal} (a).

\begin{algorithm}[t]
\caption{Dictionary Learning for Channel Representation and Precoding}
\label{Algorithm0}
\begin{algorithmic}[1]{
\renewcommand{\algorithmicrequire}{\textbf{Input:}}
\renewcommand{\algorithmicensure}{\textbf{Output:}}
\REQUIRE The intended sparsity level is $s$ ($s_c$ for channel sparsity, $s_p$ for precoding sparsity), and the number of iterations is $Num$. $\boldsymbol{X}$ is the training set composed of many training signals.} More specifically, $\boldsymbol{X}$ is a training set of channel realizations for the case of channel representation and $\boldsymbol{X}$ is DFT matrix for the case of precoding.
\ENSURE A learned dictionary $\boldsymbol{D}_H$ for channel representation and $\boldsymbol{D}_U$ for precoding.
\STATE{Obtain a signal $\boldsymbol{X}$ as
$\boldsymbol{D}^0 \gets \boldsymbol{X}$, 
initialize $i\gets 0$.}
 \WHILE{$i \leq Num$}
 \STATE{Solve:
$\argminl_ {{\boldsymbol{W}^i}} {\|\boldsymbol{X}-\boldsymbol{D}^i\boldsymbol{W}^i}\|_2^2 ~ s.t. ~ 
{\|\boldsymbol{W}^i\|}_0 <s$
 }
 \STATE{Update $\boldsymbol{D}^i$ by solving:
$\argminl_ {{\boldsymbol{D}^i}} {\|\boldsymbol{X}-\boldsymbol{D}^i\boldsymbol{W}^i}\|_2^2$
\STATE{Update: $i\gets i+1$.}
 }
 \ENDWHILE
\RETURN $\boldsymbol{D}^i$
\end{algorithmic}
\end{algorithm}

\begin{figure}[b!]
\centering
\resizebox{0.88\columnwidth}{!}{
\includegraphics[width=14cm]{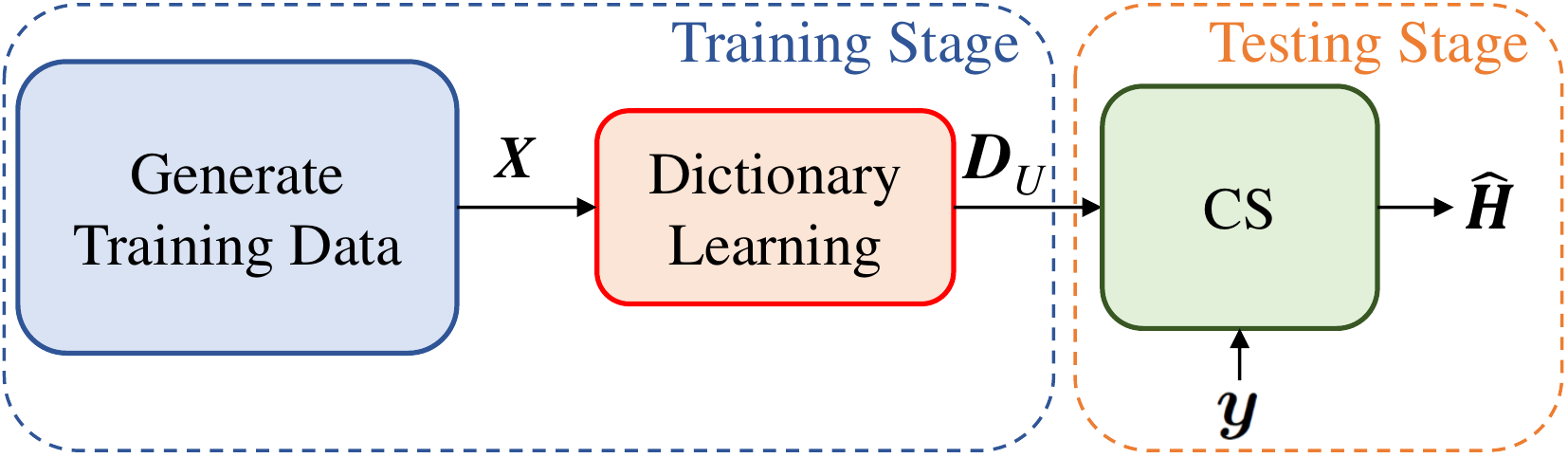}}
\caption{The diagram of the proposed dictionary learning-based algorithm for precoding.}
\label{Diagram2}
\end{figure}

\begin{figure*}[!t]
\centering
\resizebox{1.99\columnwidth}{!}{
\begin{tabular}{cc}
\includegraphics{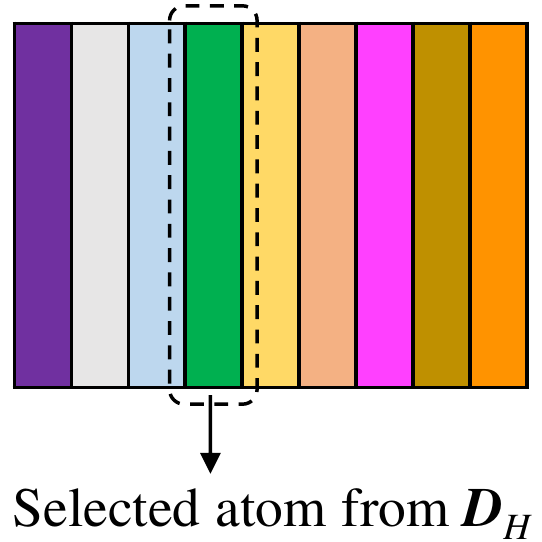}&
\includegraphics{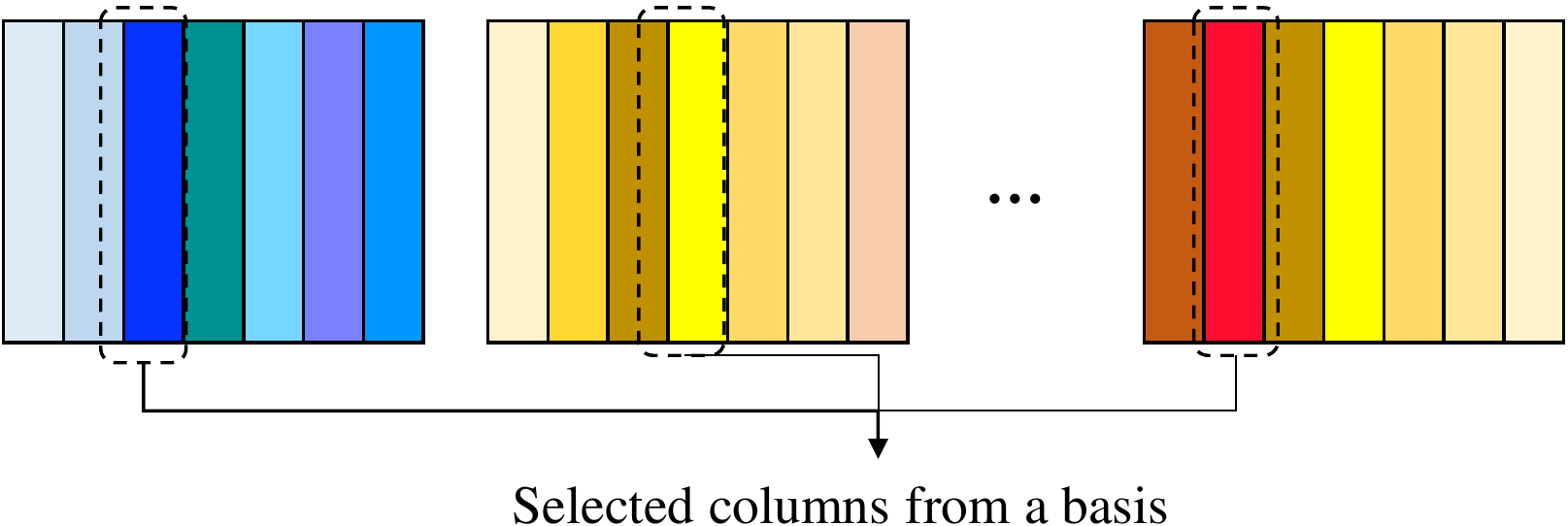} \\
\hspace{+1cm}\Huge{(a)} & \hspace{+2cm}\Huge{(b)} \\
\includegraphics{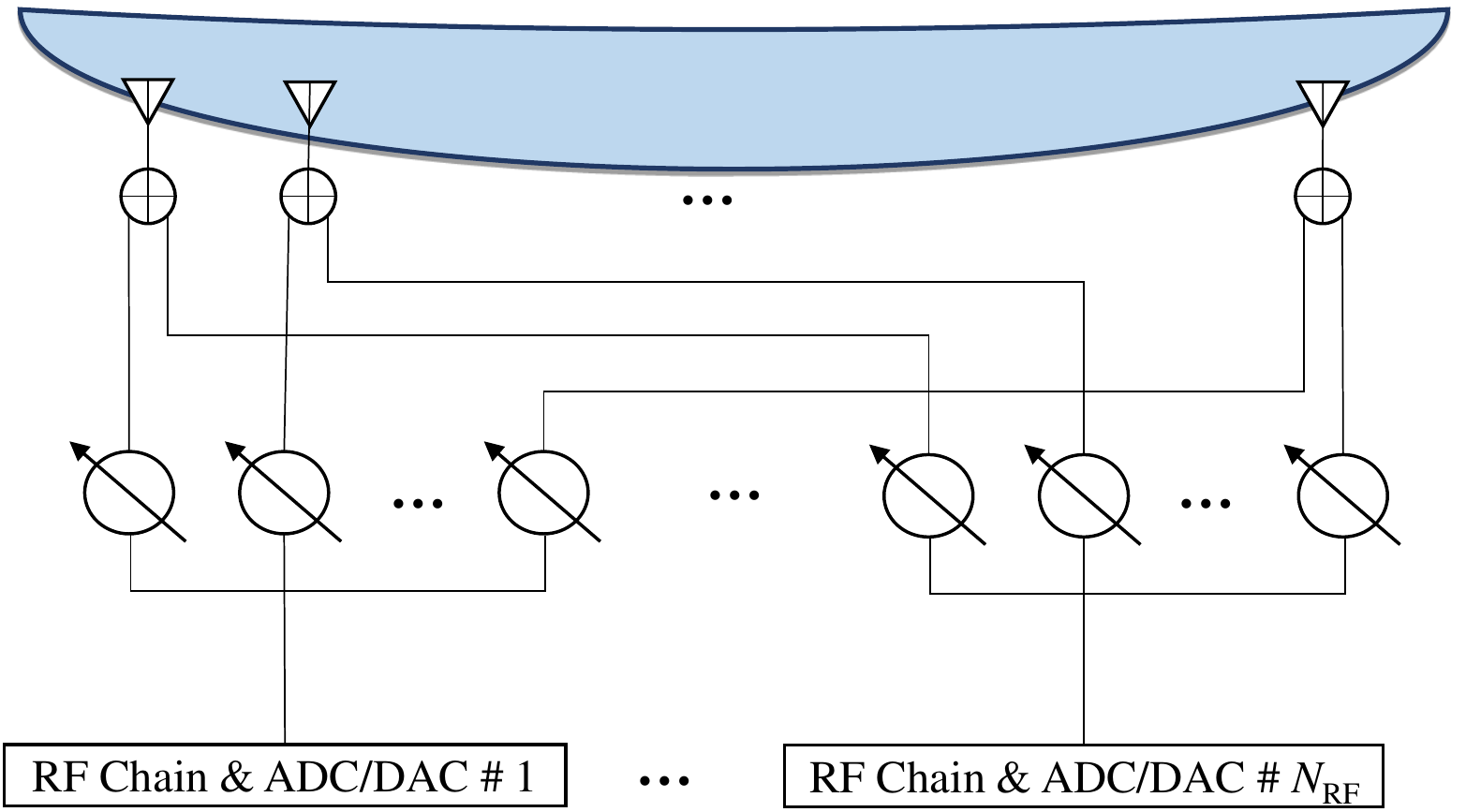}&
\includegraphics{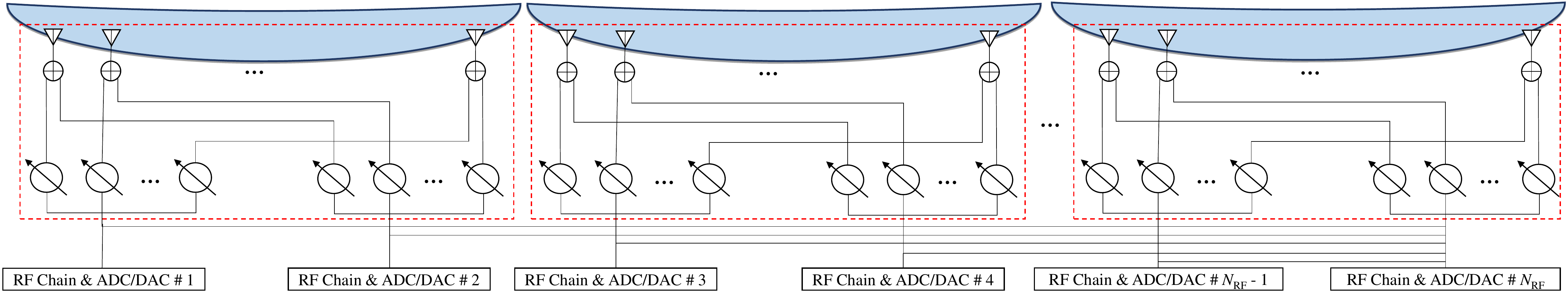}\\
\hspace{+1cm}\Huge{(c)} & \hspace{+2cm}\Huge{(d)}
\end{tabular}}
\caption{Realizing dictionary learning-based beamforming: (a) selecting a certain atom in a learned dictionary of a specific direction can be achieved by (b) selecting multiple basis functions each belonging to a certain DFT basis of a specific general directionality. Correspondingly, (c) achieving certain narrow beamforming can be realized in practice by (d) aggregating the multiple LAAs.}
\label{lens_realization}
\end{figure*}

\subsubsection{The Proposed Dictionary Learning for Precoding}

\par A block diagram of the proposed precoding through a learned dictionary is represented in Fig.~\ref{Diagram2}. This algorithm has training and testing stages as in the previous algorithm. In the training stage, a training set of DFT-based precoding matrices is generated. Afterward, a dictionary is trained over this set and a learned dictionary is created for precoding. It is noted that this process is carried out offline. Next, in the run-time, a CS algorithm is applied with the learned dictionary ($\boldsymbol{D}_{U}$) and received signal ($\boldsymbol{y}$) to estimate the beamspace channel ($\hat{\boldsymbol{H}}$). To estimate the beamspace channel with a CS algorithm, a sparse recovery algorithm in Eq. (1) is used. In this equation, $\boldsymbol{y}$ is the received signal, and $\boldsymbol{D}$ is a dictionary that is created by training the DFT precoding matrix set.

\par Let $\boldsymbol{y} \in \mathbb{C}^n$ denote a vector signal. The notion of CS considers obtaining a compressed measurement $\boldsymbol{y}_c=\boldsymbol{\Phi}\boldsymbol{y}$ where $\boldsymbol{\Phi} \in \mathbb{C}^{m\times n} $ is a measurement/sensing matrix, with $m< n$, rather than measuring every element in $\boldsymbol{y}$. Clearly, an $n$-to-$m$ dimensionality reduction is made possible by this undersampling operation. It is noted that CS is only applicable to compressible signals; those being sparse explicitly, or have a sparse representation in a certain domain \cite{davenport2010signal}. Since $\boldsymbol{y}$ is not necessarily sparse in its own shape, its sparse representation is typically obtained using a sparsifying transform/basis ($\boldsymbol{\Psi}$); either a fixed basis or a redundant (overcomplete) learned dictionary. In this context, the signal can be approximated as $\boldsymbol{y}=\boldsymbol{\Psi}\boldsymbol{w}$, where $\boldsymbol{w}$ is a sparse coding coefficient vector having only $s \ll n$ nonzero elements. Obtaining $\boldsymbol{w}$ from $\boldsymbol{y}_c$ can be formulated as follows 
\begin{equation}
\argminl_{\boldsymbol{w}} {\|\boldsymbol{w}\|}_0 ~ s.t. ~ \boldsymbol{y}_c=\boldsymbol{\Phi}\boldsymbol{y}= \boldsymbol{\Phi}\boldsymbol{\Psi}\boldsymbol{w}.\label{eq1_new}
\end{equation}

\begin{algorithm}[!t]
\caption{Beamspace Channel Representation and Precoding}
\label{Algorithm1}
\begin{algorithmic}[1]{
\renewcommand{\algorithmicrequire}{\textbf{Input:}}
\renewcommand{\algorithmicensure}{\textbf{Output:}}
\REQUIRE UL channel $\boldsymbol{H}_{UL}$, channel sparsity $s_c$, precoding sparsity $s_p$, a learned dictionary $\boldsymbol{D}_H$ for channel representation and $\boldsymbol{D}_U$ for channel estimation. 
\ENSURE A channel impulse response estimate $\hat{\boldsymbol{H}}_U$.
\STATE{Solve: 
$\boldsymbol{w}_e=\argminl_ {{\boldsymbol{w}}} {\|\boldsymbol{H}_{UL}-\boldsymbol{D}_H\boldsymbol{w}\|}_2^2 ~ s.t. ~ {\|\boldsymbol{w}\|}_0 <s_c $}
\STATE{Obtain a beamspace channel:\\
$\boldsymbol{H}=\boldsymbol{D}_H\boldsymbol {w}_e$}
\STATE{Send the signal through the $\boldsymbol{H}$.}
\STATE{Obtain $\boldsymbol{Y}$} in the receiver.
\STATE{Solve: 
$\boldsymbol {w}_u=\argminl_ {{\boldsymbol{w}}} {\|\boldsymbol{Y}-\boldsymbol{D}_{U}\boldsymbol{w}\|}_2^2 ~ s.t. ~ {\|\boldsymbol{w}\|}_0 <s_p $}
\STATE{Obtain a channel estimate:\\
$\hat{\boldsymbol{H}}=\boldsymbol{D}_{U}\boldsymbol {w}_u$}
}
\end{algorithmic}
\end{algorithm}

\par The inverse problem in (\ref{eq1_new}) is inherently ill-posed. Still, the sparsity of the solution lends itself as an efficient regularizer to this problem under mild conditions. In this regard, the restricted isometry property (RIP) \cite{Compressive_sesing} of $\boldsymbol{\Phi}$ assures a unique solution with high probability. Besides, a number of compressed measurements $m$ being at least equal to $(cs \log n/m)$ for some small constant $c > 0$ assures exact recovery according to the robust uncertainty principle \cite{Compressive_sesing}. Technically, a variety of sparse recovery techniques can be applied to obtain $\boldsymbol{w}$ given $\boldsymbol{y}_c$, $\boldsymbol{\Phi}$ and $\boldsymbol{\Psi}$. To this end, the fundamental intuition behind CS is measuring only the nonzero elements in $\boldsymbol{w}$. Hence, it resembles a compressed measurement of the original signal. Finally, the original signal can be reconstructed as $\hat{\boldsymbol{y}}=\boldsymbol{\Psi}\boldsymbol{w}$. The proposed algorithm for precoding is illustrated in Fig.~\ref{MIMOjournal} (b). Besides, the use of dictionary learning for both channel representation and precoding is shown in Fig.~\ref{MIMOjournal} (c).

\par The main steps of learning dictionaries for channel representation and precoding are detailed in Algorithm \ref{Algorithm0}. These begin by using a set of channel information (for channel representation) and a DFT matrix (for precoding) as a dictionary initialization (Step 1 of Algorithm \ref{Algorithm0}). Then, a training set is further tuned to the dictionary in both algorithms (Steps 2 through 5 of Algorithm \ref{Algorithm0}). Here note that this part is a sparse coding process where one calculates the sparse coding coefficient vectors of the given training data based on the current dictionary estimate. In other words, the algorithm implicitly approximates the solution to the $\ell_{0}$-constrained least-squares problem. The main principle behind this iterative algorithm is to use the residual error from the previous iteration to successfully approximate the position of nonzero entries and estimate their values. More details of this part can be found in \cite{bahmani2013greedy}.

\par All testing stages of the proposed beamspace channel representation and precoding algorithms are detailed in Algorithm \ref{Algorithm1}. In this stage, a sparse coding vector is obtained according to the UL channel and the learned dictionary of channel representation (Step 1 of Algorithm \ref{Algorithm1}). Then, the beamspace channel is obtained according to this sparse coding vector and the learned dictionary of the channel representation (Step 2 of Algorithm \ref{Algorithm1}). Afterward, the signal is sent through this beamspace channel and the signal is received by the receiver (Steps 3 and 4 of Algorithm \ref{Algorithm1}). Finally, the channel is estimated according to the learned dictionary of channel estimation and received signal (Steps 5 and 6 of Algorithm \ref{Algorithm1}). 

\vspace{-0.2cm}

\subsection{Realization of Beamspace MmWave MMIMO Hardware by A Learned Dictionary}

\par Implementing sparse coding over a dictionary requires a particular type of an LAA that synthesizes the sparse coding over the dictionary atoms. For this purpose, one can employ a set of classical LAAs. In this setting, the aggregate effect of the composite LAAs would translate to the intended sparse coding over the dictionary. This setting is motivated by the idea that a dictionary atom can be expanded in terms of several DFT basis vectors as in the Appendix. Therefore, the dictionary, as a whole, can be cast as a composition of multiple DFT transformations of different orientations. Thus, dictionary learning can be configured with the traditional function of beam selection.

\par Analogous to the way a dictionary selectively chooses specific basis vectors (called atoms) to represent such a signal, one can use a phase shifter network to selectively choose certain phase shifters to create a specific beam. In other words, beam selection attained by controlling the switches in this network mimics atom selection in a given dictionary based on the atom’s similarity to the signal of interest. Several phase shifter networks can be combined to realize dictionary learning. In these networks, some phase shifters are turned off to realize “unselect” \cite{chen2012terahertz} and set some phase shifters to shift the phase 0 degree to realize “select” in beam selection. Alternatively, the adaptive selecting network \cite{Dai_Journal} can be directly utilized to design an analog precoder for data transmission, which can further improve the performance. One possible way is to extend the simple conjugate analog precoder \cite{liang2014low} to the scenarios where only 1-bit phase shifters are used. More efficient schemes will be left for our further work.

\par The realizations of the algorithm in \cite{Dai_Journal} and the proposed algorithm are illustrated in Fig.~\ref{lens_realization} with their atom selection strategies. Here note that, in the illustration, phase shifter network is obtained with 1-bit phase shifters as in \cite{Dai_Journal}. Still, this may also be a little complicated. If a specific performance loss is permitted, the phase shifters can be also replaced with switches, and the proposed scheme can be employed.

\subsection{Discussion on Computational Complexity}

\par The computational complexity of the proposed algorithm mainly depends on sparse coding and dictionary learning. Let us consider the naive OMP algorithm as an example of sparse coding, where it is working on sparse coding of a signal $\boldsymbol{x} \in \mathbb{C}^N$ over a given dictionary $\boldsymbol{D} \in \mathbb{C}^{N \times K}$. Its computational complexity at the $k$th iteration is $\bigO(NK+Ks+Ks^2+s^3)$ \cite{sturm2012comparison}. With sparsity $s$, the overall complexity of the OMP algorithm is $\bigO(NKs+Ns^2+Ns^3+s^4)$. Note that sparse coding is used both during the training and testing stages. The K-SVD \cite{ksvd} algorithm can be considered as an example for dictionary learning process. The total complexity of K-SVD working on a training set $\boldsymbol{X}\in \mathbb{C}^{N \times L}$, with sparsity $s$ and $Num$ iterations is $\bigO(Num(s^2+N)KL)$ \cite{skretting2010recursive}. It is noted, dictionary learning is performed only in the training stage.

\section{Simulations and Results}
\label{Section4}

\subsection{Parameter Setting}

\par This paper considers a mmWave mMIMO system with $N=$ 256 antennas and $N_{\text{RF}}=$ 16 RF chains. This system simultaneously serves 16 UEs at the receiver end. With the SV channel model, similar to the experimental setup in \cite{Dai_conf}, the $k$th UE spatial channel is obtained as a composition of one LoS component and two NLoS components. These are set to have $\beta_k^{(0)}\sim \mathcal{CN} (0, 1)$ and $\beta_k^{(i)} \sim \mathcal{CN} (0, {10}^{-0.5})$ for $i=$ 1, 2; 3. $\psi_k^{(0)}$ and $\psi_k^{(i)}$ follow the independent and identically distributed (i.i.d.) uniform distribution within $\psi \in [-0.5,0.5]$.

\par For simulating the GSCM, the experimental setup used in \cite{Ding_Rao} is used. This setup considers a system made up of a single urban cell of a radius of 1200 meters, with the BS at its center. The DL channel is generated according to the GSCM principles \cite{GSCM} with coefficients provided by the spatial channel model \cite{3gpp_report}. Also, the azimuth angle $\theta$ ranges between $-\pi/2$ and $\pi/2$. As for the scattering environment, the cell has seven fixed-location scattering clusters. The distance between each cluster and the BS is selected randomly in ranges between 300 meters and 800 meters. Four scattering clusters are used for each channel modeling; one is at the UE location, the remaining three clusters are the closest to the UE from the previously mentioned seven scattering clusters. The UE location is spanned consistently to be between 500 meters and 1200 meters. Under the GSCM guidelines, each scattering cluster has 20 effective propagation subpaths with a 4-degree angular spread.

\begin{figure}[t]
\centering
\resizebox{0.77\columnwidth}{!}{
\includegraphics[width=14cm]{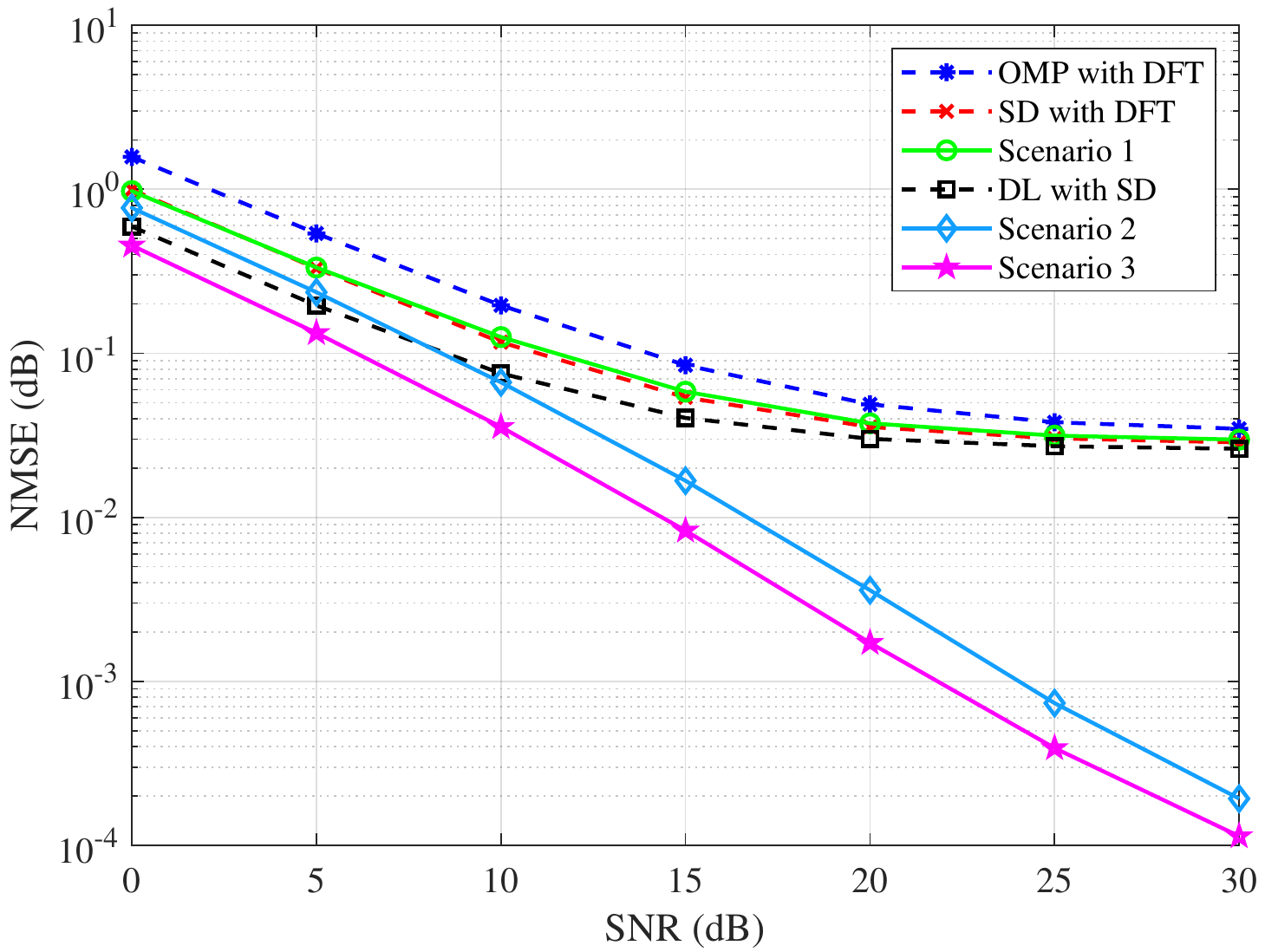}}
\caption{ULA NMSE performance comparison versus SNR with the SV channel model.}
\label{NMSESV}
\end{figure}

\begin{figure}[t]
\centering
\resizebox{0.77\columnwidth}{!}{
\includegraphics[width=14cm]{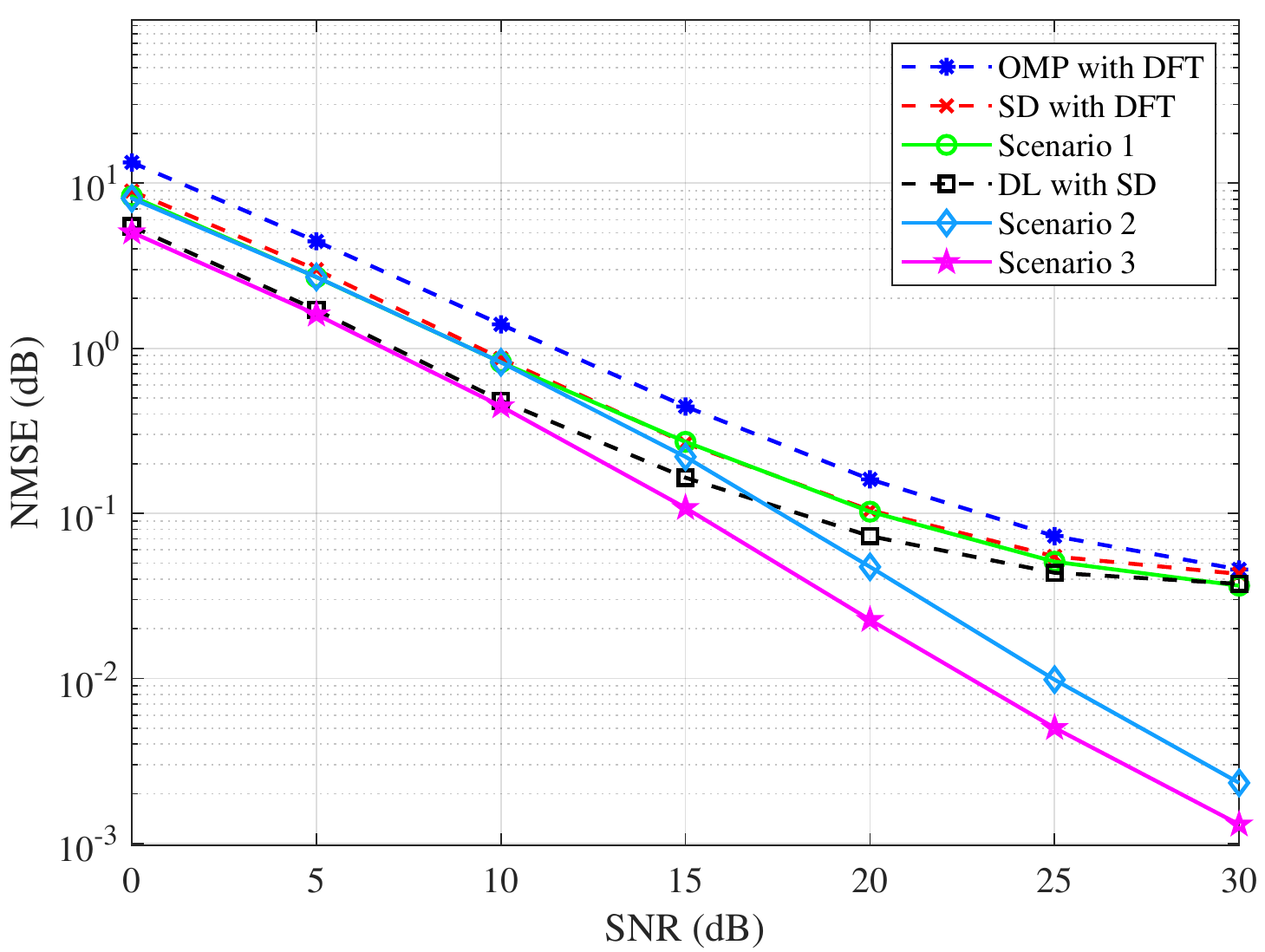}}
\caption{ULA NMSE performance comparison versus SNR values with the GSCM.}
\label{NMSEGSCM}
\end{figure}

\begin{figure}[t]
\centering
\resizebox{0.77\columnwidth}{!}{
\includegraphics[width=14cm]{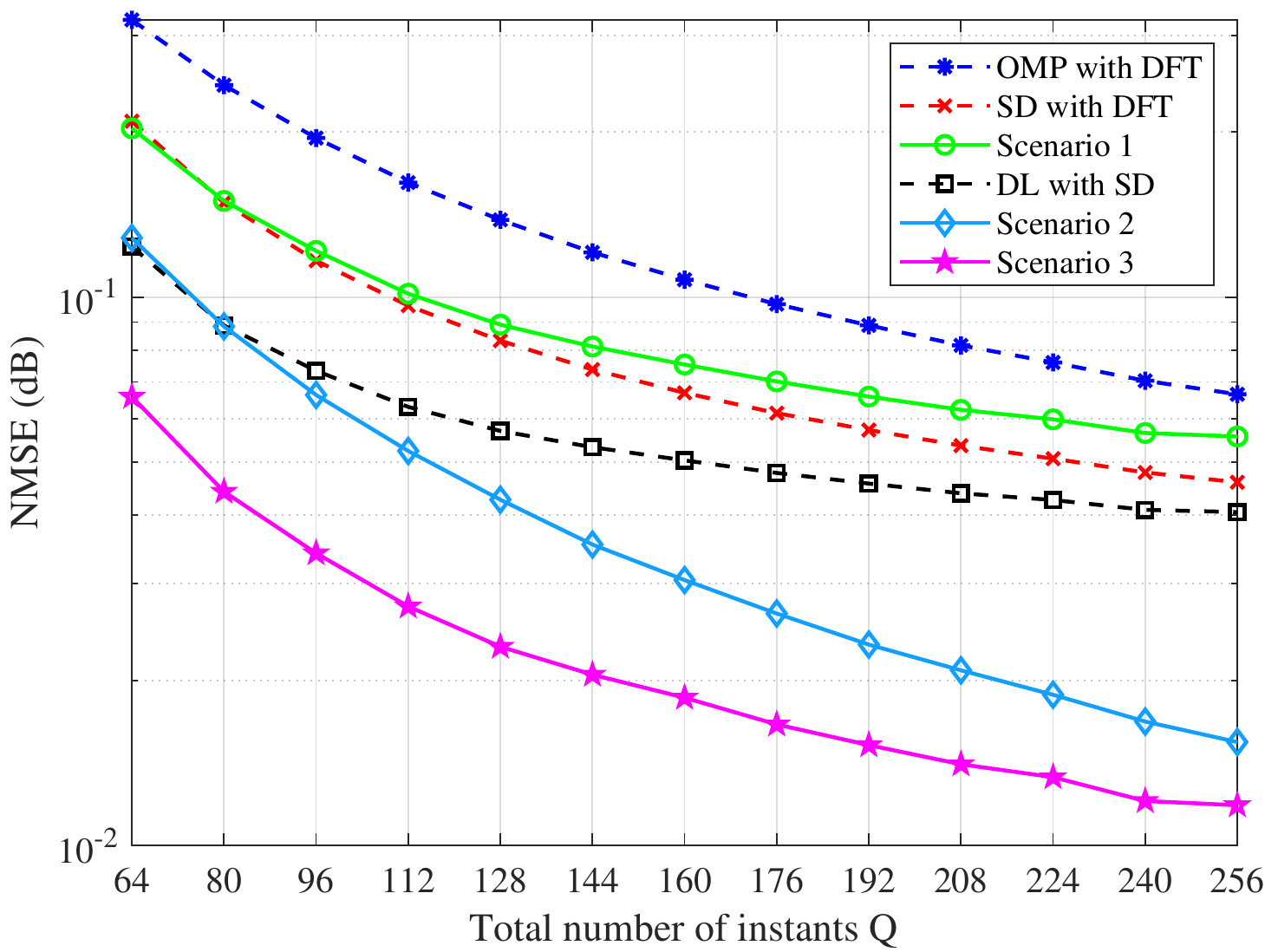}}
\caption{ULA NMSE performance comparison against the total number of instants $Q$ for pilot transmission in SV channel model.}
\label{NMSEQSV}
\end{figure}

\begin{figure}[t]
\centering
\resizebox{0.77\columnwidth}{!}{
\includegraphics[width=14cm]{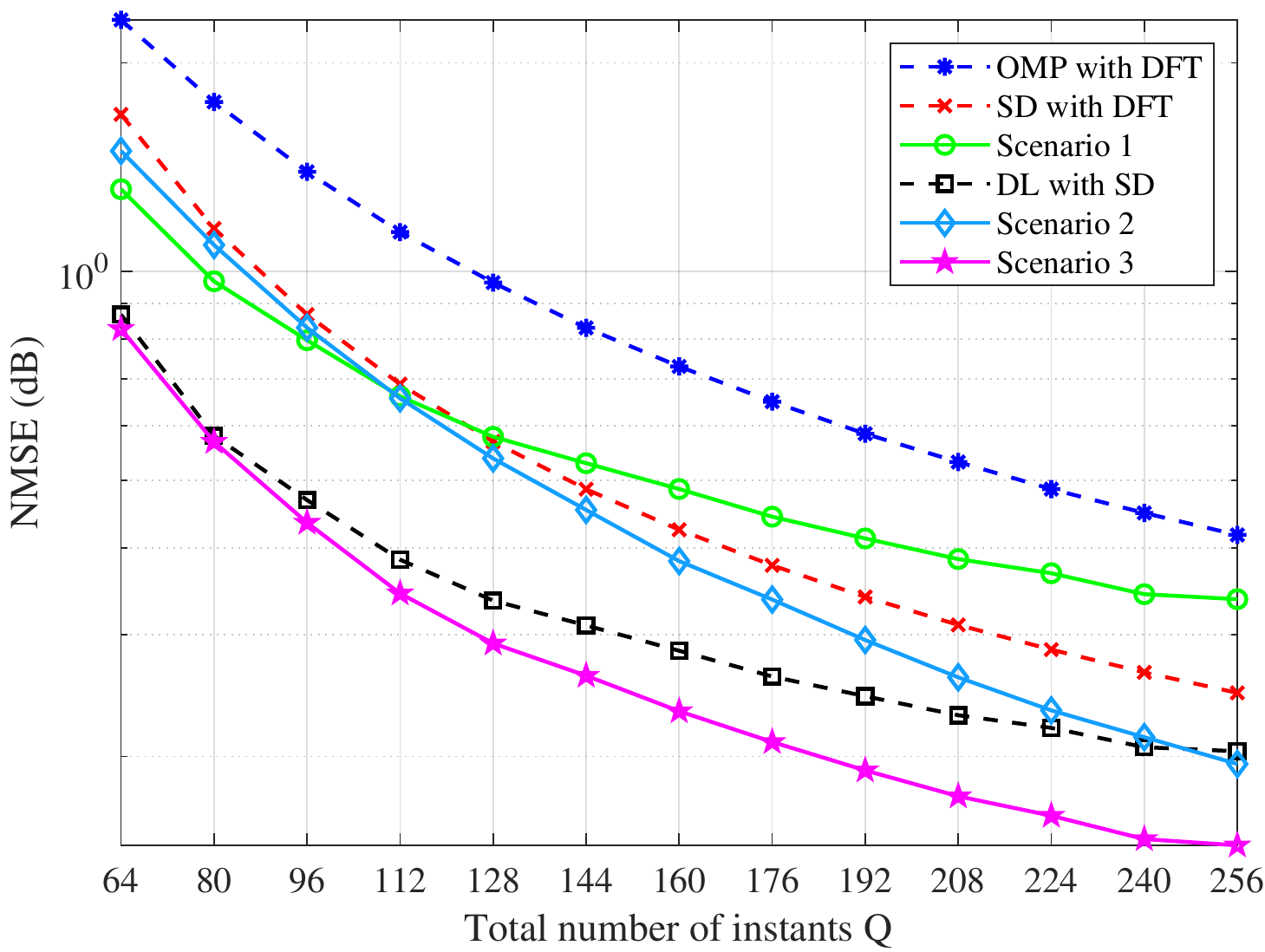}}
\caption{ULA NMSE performance comparison against the total number of instants $Q$ for pilot transmission in GSCM.}
\label{NMSEQGSCM}
\end{figure}

\par For dictionary learning, we use a training set of 10000 training vectors using the K-SVD algorithm \cite{ksvd} with 50 iterations, and a sparsity level of 16. All the experiments are made with 5000 trials. Also, it is assumed that the true values of the channel realizations are known by the transmitter in the dictionary learning stage for the sake of simplicity.

\subsection{Performance Evaluation}

\par The channel estimation performance is evaluated in terms of the normalized mean-square error (NMSE) quality metric. Then, the sum-rate performance is considered as a secondary quality metric. In this text, we compare the following algorithms.

\begin{itemize}[leftmargin=*]
\item $OMP\; with\; DFT$: OMP channel estimation when DFT bases are used for channel representation and precoding 
\item $SD\; with\; DFT$: SD algorithm (\cite{Dai_conf, Dai_Journal} where OMP-based estimation is followed by a least-squares update exploiting the structure of mmWave mMIMO channels in beamspace) when DFT bases are used for channel representation and precoding 
\item $Scenario\; 1$: OMP channel estimation when a DFT basis and a learned dictionary are used for channel representation and precoding, respectively
\item $DL\; with\; SD$: SD algorithm when a DFT basis and a learned dictionary are used for channel representation and precoding, respectively
\item $Scenario\; 2$: OMP channel estimation when a learned dictionary and DFT basis are used for channel representation and precoding, respectively
\item $Scenario\; 3$: OMP channel estimation when learned dictionaries are used for channel representation and precoding
\end{itemize}

\par First, the NMSE performance of the aforementioned channel representation and estimation settings versus SNR is investigated. This experiment is first performed with the SV channel model and then with the GSCM. A ULA is considered for both models. The results of these settings are shown in Figs.~\ref{NMSESV} and \ref{NMSEGSCM}, respectively. For SD-based channel estimation, we keep the strongest $V=$ 9 elements for each channel component and assume that the sparsity level of the beamspace channel for the OMP-based channel estimation is equal to $V(L+1)=$ 16. We also assume that all channel estimation algorithms use $Q=$ 96, training pilots.

\par In view of Figs.~\ref{NMSESV} and~\ref{NMSEGSCM}, it is evident that using a learned dictionary in the precoding improves the channel estimation quality. This is the case for both OMP-based reconstruction, and the SD algorithm. Also, using a learned dictionary channel representation further improves the performance, especially for high SNR values.

\par Next, the previous experiment is repeated with the difference that SNR is fixed at 10 $dB$ and the number of training pilots ($Q$) is varied. The results are depicted in Figs.~\ref{NMSEQSV} and \ref{NMSEQGSCM} for the SV channel model and GSCM, respectively. In view of these figures, it is shown that for the same $Q$, using learned dictionaries for channel representation and precoding improves the NMSE performance. Said equivalently, using a learned dictionary allows for reducing the training overhead for having the same NMSE performance attained with a DFT basis.

\begin{figure}[t]
\centering
\resizebox{0.77\columnwidth}{!}{
\includegraphics[width=14cm]{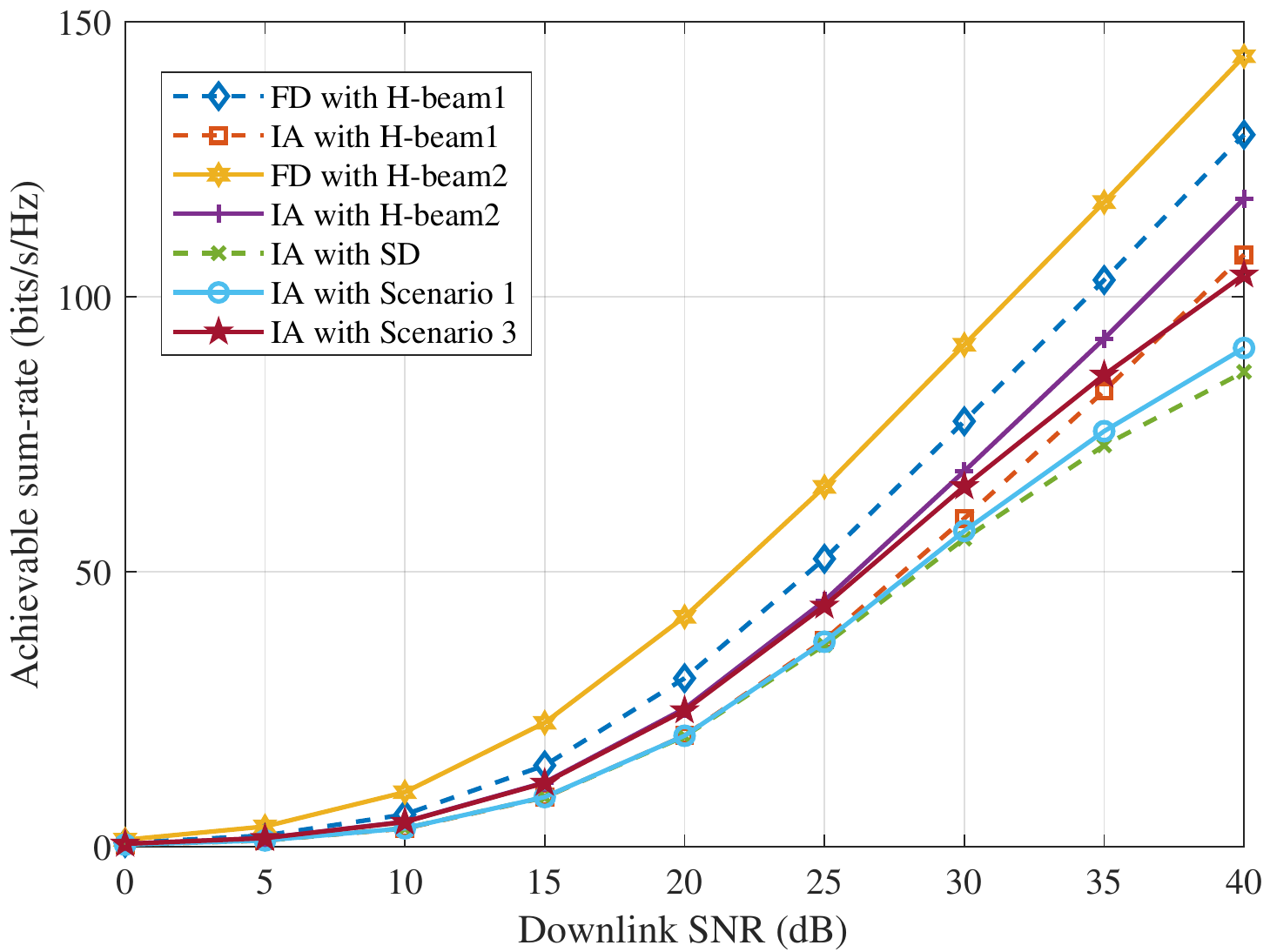}}
\caption{ULA sum-rate comparison for DFT and dictionary learning-based algorithms.}
\label{IA}
\end{figure}

\begin{figure}[t]
\centering
\resizebox{0.77\columnwidth}{!}{
\includegraphics[width=14cm]{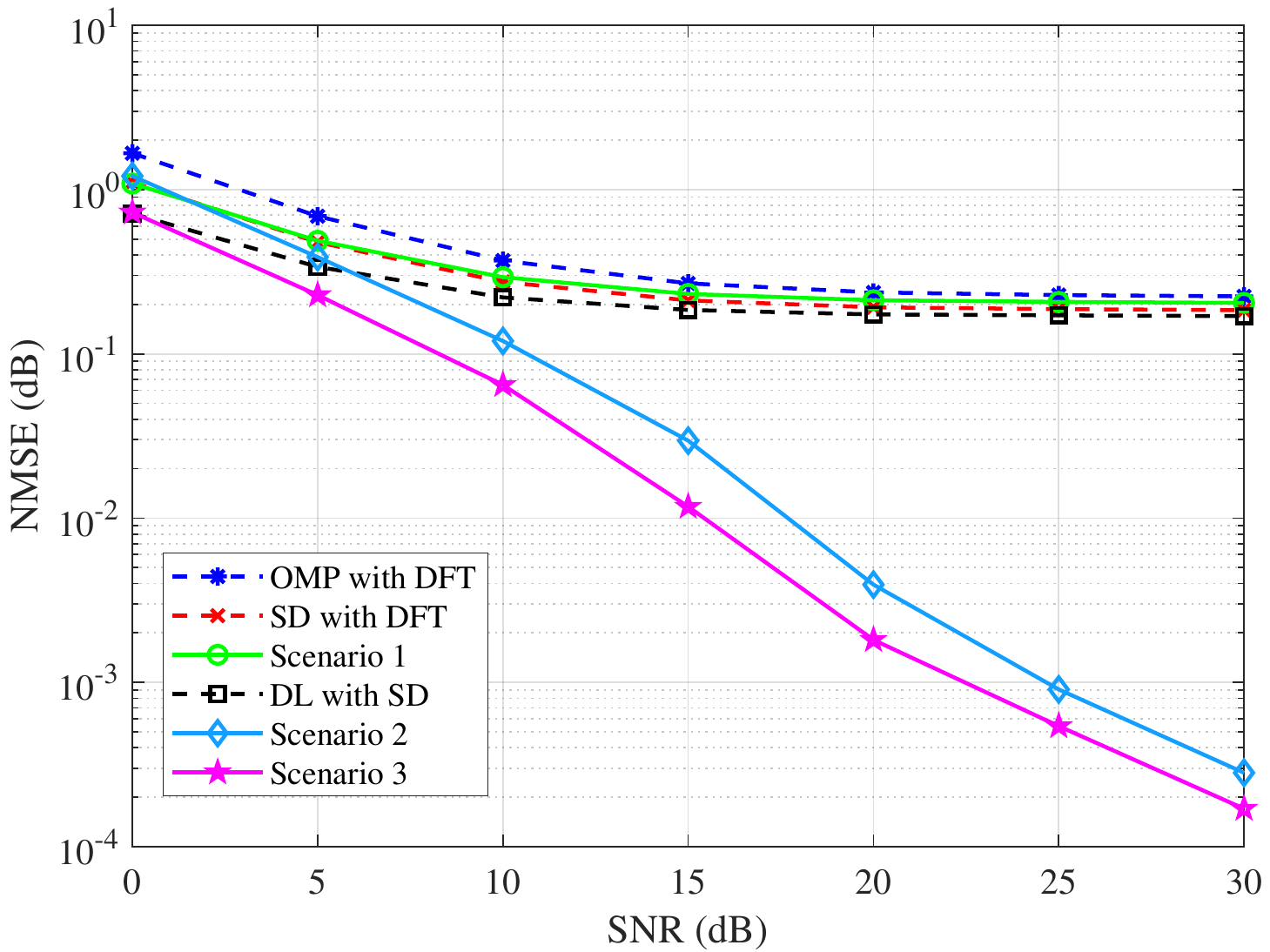}}
\caption{NULA NMSE performance comparison against different SNR values for SV channel model.}
\label{NMSEnonSV}
\end{figure}

\begin{figure}[t]
\centering
\resizebox{0.77\columnwidth}{!}{
\includegraphics[width=14cm]{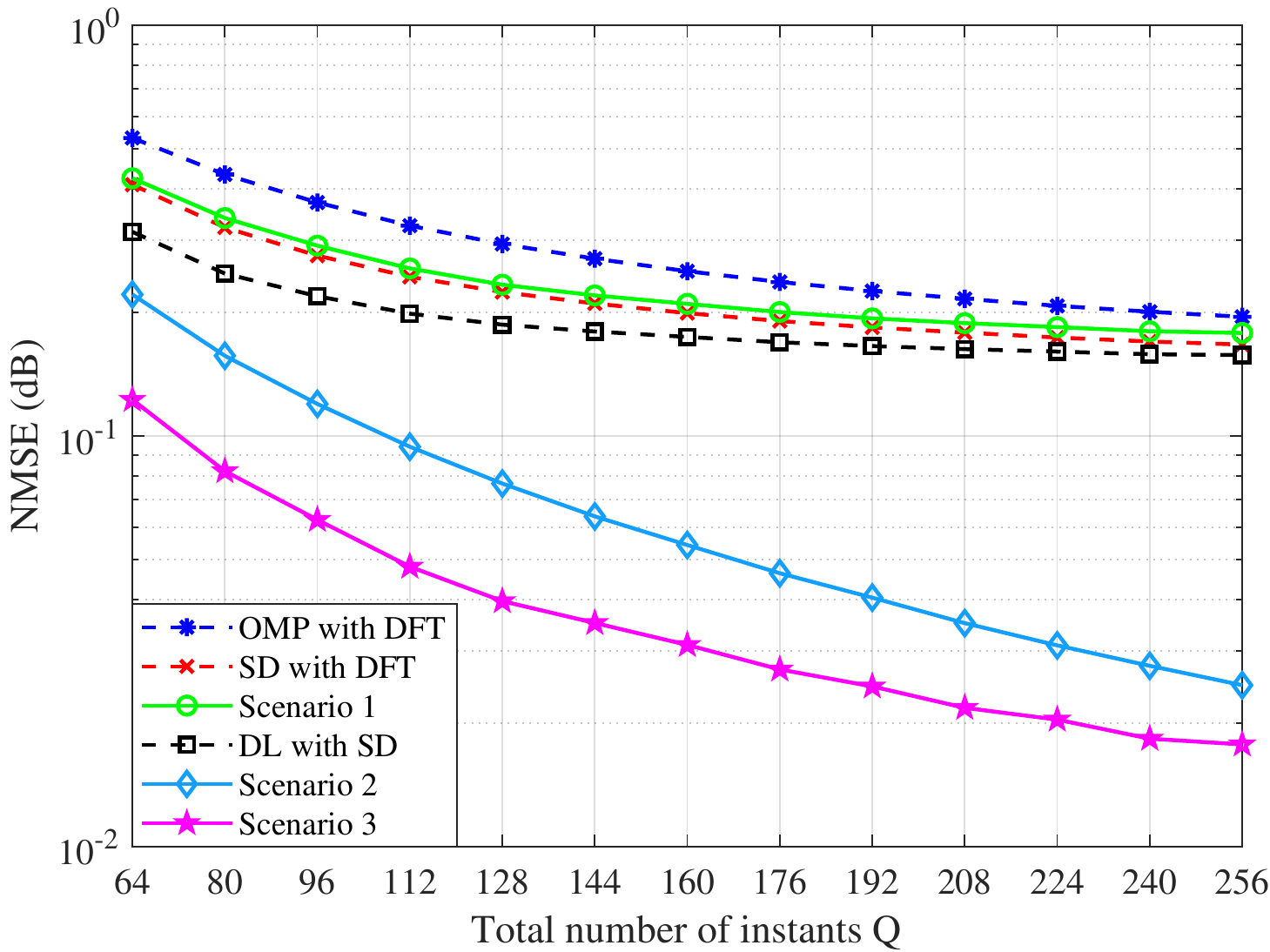}}
\caption{NULA NMSE performance comparison against the total number of instants $Q$ for pilot transmission for SV channel model.}
\label{nonlinearNMSEQSV}
\end{figure}

\begin{figure}[t]
\centering
\resizebox{0.77\columnwidth}{!}{
\includegraphics[width=14cm]{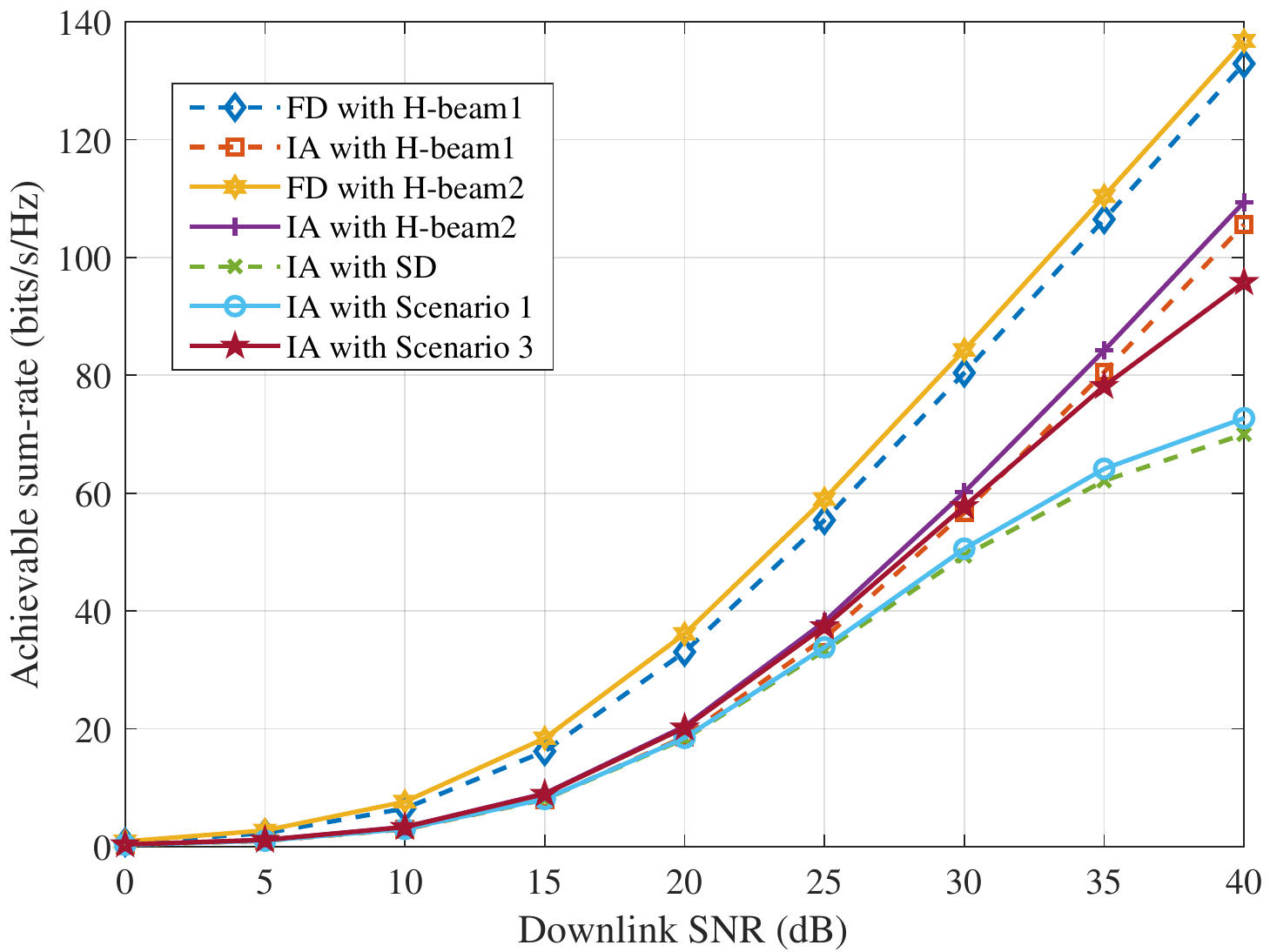}}
\caption{NULA sum-rate comparison for DFT and dictionary learning-based algorithms.} 
\label{IAnonlineer}
\end{figure} 

\par The quality of channel estimation is measured in terms of beams selection. The following scenarios are compared for this purpose. 
\begin{itemize}[leftmargin=*]
\item $FD$: Fully digital zero-forcing (ZF) precoders, included as a benchmark when a DFT basis ($FD\; with\; H$-$beam1$) and a learned dictionary ($FD\; with\; H$-$beam2$) are used for channel representations
\item $IA$: Interference-aware (IA) beam selection algorithm \cite{IA} which assume perfect beamspace channel knowledge when a DFT basis ($IA\; with\; H$-$beam1$) and a learned dictionary ($IA\; with\; H$-$beam2$) are used for channel representations
\item $IA\; with\; SD$: IA fed with a beamspace channel estimate obtained with SD when DFTs are used for precoding and channel representation
\item $IA\; with\; Scenario\; 1$: IA fed with a beamspace channel estimate obtained with SD when a DFT basis and a learned dictionary are used for channel representation and precoding, respectively
\item $IA\; with\; Scenario\; 3$: IA fed with a beamspace channel estimate obtained with OMP when learned dictionaries are used for precoding and channel representation
\end{itemize}
\noindent Here, the previously mentioned parameter setting is used. The results are shown in Fig.~\ref{IA}.

\par As clearly seen in Fig.~\ref{IA}, fully digital ZF algorithms achieve the best sum rate. Next, are the IA algorithms with perfect beamspace channel knowledge. In both cases, dictionary learning-based algorithms are superior to DFT-based algorithms. This verifies the improvement of the channel representation quality when a learned dictionary is used. Also, the performance of the proposed $IA\; with\; Scenario\; 3$ algorithm is very close to the perfect IAs. Besides, the IA algorithm fed with the beamspace channel estimate obtained with a learned dictionary is consistently better than the case of the SD algorithm.

\par Finally, all the simulations are done with a non-ULA (NULA)\footnote{NULA case defines the manufacturing error and evaluates the irregular array geometries is made by assuming that the antenna spacing is uniformly distributed within 0.45$\lambda$ and 0.55$\lambda$, where $\lambda$ is the carrier wavelength.}. Here, we provide only simulations with the SV channel model, to avoid repetition. For the GSCM, similar behavior in the graphs is observed. Figures.~\ref{NMSEnonSV}, ~\ref{nonlinearNMSEQSV}, and ~\ref{IAnonlineer} show that the behaviors are similar to the case of using a ULA. However, the advantages of using learned dictionaries are more strongly pronounced in the NULA case. This is especially the case with high SNR values. However, in the low SNR regime, the improvement is not significant.

\section{Conclusions}
\label{Section5}

\par This paper proposed the use of learned dictionaries as the sparsifying transform operators used in creating beamspace channels in mmWave mMIMO. This corresponds to the use of composite LAAs that enhance the beamspace sparsity. This enhancement leads to a more efficient pilot reduction in comparison to the standard case of using LAAs corresponding to fixed basis functions. Dictionary atoms have been shown to possess riches structures compared to DFT basis functions. A learned dictionary has been shown to reduce the phenomenon of power leakage in mmWave mMIMO due to the uses of such atoms. Similarly, we proposed the use of a learned dictionary to function as the precoding operator matrix, meeting the same objective of channel sparsity enhancement. Numerical experiments have shown that these contributions lead to improving the quality of channel estimation and spectral efficiency, as validated in terms so the NMSE and sum-rate performance measure. It is noted that the performance improvement is especially strong in the cases on a NULA. As future work, the proposed algorithm will be investigated with more practical scenarios, e.g., a planar array with azimuth and elevation, and broadband communications.

\appendix

\par Power leakage can be viewed as an imperfection in the sparse representation obtained with a given sparsifying basis. Thus, we compare the quality of a sparse representation over a DFT basis $\boldsymbol{F} \in \mathbb{C}^{N \times N}$ to that over a redundant (overcomplete) dictionary $\boldsymbol{D} \in \mathbb{C}^{N \times K}$, where $K> N$. In this setting, the signal of interest is a (mmWave mMIMO) beamspace channel $\boldsymbol{h} \in \mathbb{C}^{N}$. Now, let us compare these representations with a sparsity level $s$.

\par First, an exact representation of $\boldsymbol{h}$ over $\boldsymbol{F}$ can be obtained using the whole $N$ basis functions (columns) in $\boldsymbol{F}$, as follows
\begin{equation}
 \boldsymbol{h}= \boldsymbol{F}_1 a_1 + \boldsymbol{F}_2 a_2 + \dots + \boldsymbol{F}_N a_N,
\end{equation}
\noindent where $a_1$ through $a_N$ denote the representation coefficients of $\boldsymbol{h}$ with respect to $\boldsymbol{F}$. These can be obtained by performing an inner product between $\boldsymbol{h}$ and $\boldsymbol{F}$.

\par An $s$-sparse representation of $\boldsymbol{h}$ over $\boldsymbol{F}$ can be obtained by selecting the most dominant $s$ coefficients. For simplicity, let us assume that they happen to be the first $s$ coefficients, as follows
\begin{equation}
 \boldsymbol{\hat{h}}_F= \boldsymbol{F}_1 a_1 + \boldsymbol{F}_2 a_2 + \dots + \boldsymbol{F}_s a_s.
\end{equation}

\par Second, with respect to $\boldsymbol{D}$, an $s$-sparse representation of $\boldsymbol{h}$ is
\begin{equation}
 \boldsymbol{\hat{h}}_D= \boldsymbol{D}_1 b_1 + \boldsymbol{D}_2 b_2 + \dots + \boldsymbol{D}_s b_s.\label{eq100}
\end{equation}
\noindent Again for simplicity, let us assume that the first $s$ atoms (columns) of $\boldsymbol{D}$ are selected, with the corresponding coefficients $b_1$ through $b_s$.

\par Each dictionary atom is a prototype signal that is rich in structure, as shown in the motivating example of Section \ref{Section3}-B. Thus, one can assume that it can be expanded spanning many DFT basis functions. So, it can be written as:
\begin{equation}
 \boldsymbol{D}_1= \boldsymbol{F}_1 c_1 + \boldsymbol{F}_2 c_2 + \dots + \boldsymbol{F}_{K} c_{K}.
\end{equation}
\noindent where $K$ is the number of DFT columns required to represent the dictionary atom $\boldsymbol{D}_1$ with coefficients $c_1$ through $c_{K}$. Similarly, the atoms $\boldsymbol{D}_2$ through $\boldsymbol{D}_s$ can be expanded using $K+1$ through $K+s-1$ columns from $F$.

\par Now, (\ref{eq100}) can be rewritten as follows
\begin{equation}
\centering
\begin{split}
\boldsymbol{\hat{h}}_D=& (\boldsymbol{F}_1 c_1 + \dots+ \boldsymbol{F}_{K} c_{K}) b_1 +\\
& \dots +\\
& (\boldsymbol{F}_1 d_1 + \dots+ \boldsymbol{F}_{K} d_{K}) b_s. \\
\end{split}
\end{equation}
\par From the last formulation, it is evident that using the same sparsity level, the sparse representation of $\boldsymbol{h}$ over $\boldsymbol{D}$ is $s$-sparse, in terms of sparsity. However, it is richer in terms of the structure as it is equivalent to using many columns from $\boldsymbol{F}$ \cite{starck2015sparse}. Said conversely, one can obtain a sparser representation over $\boldsymbol{D}$ with almost the same representation quality.

\end{document}